# Quantifying Personality in Human-Drone Interactions for Building Heat Loss Inspection with Virtual Reality Training

Pengkun Liu [a]*, Pingbo Tang [a], Jiepeng Liu[b], Yu Hou[c]*

[a] *Department of Civil and Environmental Engineering, Carnegie Mellon University, 5000 Forbes Avenue, Pittsburgh, PA, 15213, United States*
[b] *School of Civil Engineering, Chongqing University, Chongqing 400045, China*
[c] *Department of Construction Management, Western New England University, 1215 Wilbraham Rd, Springfield, MA 01119, United States*

* Corresponding Author

## Abstract

Reliable building energy audits are crucial for reducing waste and improving energy efficiency, particularly through the detection of heat loss in building envelopes. While sensor-equipped drones and AI-powered solutions assist in path planning and refining human operators' control actions, they often overlook the nuanced interplay between personality traits, stress management, and operational strategies that expert engineers employ when adjusting flight paths based on real-time observations. This gap underscores the challenge of achieving accurate and efficient drone-based inspections, necessitating both methodological adjustments and a comprehensive understanding of human cognitive and behavioral factors. Moreover, workforce shortages due to retiring experts necessitate effective knowledge transfer to train the next generation of inspectors. This study proposes a virtual reality (VR)-based workforce training system designed to improve human-drone interaction for building heat loss inspection. Participants piloted a virtual drone equipped with a thermographic monitor to identify defects in a simulated environment. By analyzing flight trajectory patterns, stress adaptation, and inspection performance across trainees with diverse engineering backgrounds and personality traits, we uncovered key insights: 1. Flight Trajectory Patterns – Extraverts, Intuitives, Feelers, and Perceivers explored larger areas but exhibited higher misclassification rates, while Introverts, Sensors, Thinkers, and Judgers demonstrated methodical, structured approaches. 2. Stress Adaptation – Analysis of heart rate variability (HRV) revealed broader stress fluctuations among Extraverts, Intuitives, Feelers, and Perceivers, whereas Introverts, Sensors, Thinkers, and Judgers maintained steadier physiological responses under demanding tasks. Task complexity further magnified these differences, influencing performance under pressure. 3. Inspection Performance –Extraverts, Intuitives, and Feelers achieved higher recall and coverage but were prone to over-identifying non-defective areas. Conversely, Introverts, Sensors, Thinkers, and Judgers made fewer random errors but risked overlooking subtle heat losses. These insights highlight the interplay among personality traits, stress management, and operational strategies in VR-based training for drone-assisted building audits. The proposed framework shows potential for addressing workforce shortages by facilitating tacit knowledge transfer and optimising human–drone collaboration. This study advances adaptive training paradigms for the evolving demands of intelligent building diagnostics.

## 1. Introduction

Auditing building energy usage and inspecting energy infrastructure are critical for reducing waste and improving overall energy efficiency. As cities move toward decarbonisation and the development of smart technologies, building energy audits and inspections require careful planning, specialized training, and advanced equipment to ensure both the safety of inspectors and the reliability of the machinery. Recent advancements in sensor-equipped unmanned aerial vehicles (UAVs), commonly known as drones (equipped with optical, thermal, and hyperspectral cameras), have enhanced the intelligence and efficiency of building inspections, particularly for heat loss detection in building envelopes. Despite these technological improvements, drone-assisted building audits pose complex challenges that extend beyond technical capabilities to human factors influencing inspection quality.

While AI-powered solutions assist in path planning and refining human operators' control actions, they often overlook the nuanced interplay between personality traits, stress management, and operational strategies that expert engineers employ when adjusting flight paths based on real-time observations. Operators must not only possess strong drone piloting skills but also optimize flight paths in real-time, maintain situational awareness, and make rapid decisions about potential defects. Additionally, building inspection using camera-mounted drones involves issues of human-robot trust and effective human-robot interaction (HRI). Most critically, successful drone-assisted inspections depend on experienced engineers who can quickly identify common areas of heat loss based on their knowledge of building physics and minimize rework through effective inspection strategies. These complex skills involve tacit knowledge that is notoriously difficult to capture and formalize, creating a significant knowledge transfer bottleneck.

This knowledge transfer challenge is especially urgent given current workforce shortages due to retiring experts, necessitating effective methods to train the next generation of inspectors. Further complicating training efforts are individual differences in cognitive processing styles and personality traits—dimensions often overlooked in conventional training programs. While virtual reality (VR) simulations show promise for skill acquisition, their design frequently neglects neuroadaptive principles that align instructional scaffolding with learners' personality traits. The Myers-Briggs Type Indicator (MBTI) framework offers a structured lens to examine how different affect thermal pattern recognition [1].

Against this backdrop, we propose a physiologically instrumented VR learning environment prototype tailored for training in drone-assisted building heat loss inspection that considers users' MBTI profiles. This research involves the acquisition of physiological data as participants navigate simulated inspection scenarios, along with collecting their feedback through self-reported measures. Our aim is twofold: (1) to implement and test this prototype among engineering students by analyzing their behaviors based on flight trajectory, heart rate variability (HRV), and heat loss detection performance, and (2) to examine performance differences across personality dimensions and explain the underlying mechanisms. Through this study, we address two primary research questions:

(1) What multi-dimensional metrics (behavioral, physiological, operational) optimally quantify participants' performance in a VR learning environment for building heat loss inspection?

(2) How do participants' personality dimensions impact their flight trajectory patterns, stress adaptation, and inspection performance in drone-assisted building audits?

Based on our observations of participants with diverse backgrounds, we test several hypotheses:

(1) **HP1:** Students' physiological responses exhibit noticeable variations based on the experiments encountered.

(2) **HP2**: Participants' performance and drone operation preferences vary based on their diverse backgrounds, the personalities categorized by the MBTI test, and the complexity level of the tasks.

(3) **HP3:** Participants' stress levels change regarding the content of the tasks. Their stress level reduces when they are used to the virtual environments.

The rest of this paper is organized as follows. Section 1 introduces the research background, motivation, objectives, and potential contributions. Section 2 reviews the literature on building energy audits, human-robot interaction, and simulation-based learning, identifying key research gaps and challenges. Section 3 describes the design and implementation of our immersive learning environment and the research methods, study design, including details of the user interface, drone controls, data visualization, and feedback mechanisms. Section 4 presents the experimental evaluation including data collection and analysis, and the resulting findings. Finally, Section 5 summarizes the main contributions, discusses implications, and offers directions for future research.

## 2. Literature Review

### 2.1 Drone-Assisted Building Energy Audits

Building energy audits are systematic processes of assessing buildings' energy performance and efficiency to identify the sources of energy consumption, waste, and loss and the potential opportunities for energy saving and improvement [2]. Building energy audits can also provide recommendations and solutions for enhancing energy efficiency and reducing the greenhouse gas emissions of buildings. The audits can be classified into different levels according to the depth and scope of the analysis, ranging from preliminary or walk-through audits to detailed or comprehensive audits. For building inspection, various onsite non-destructive evaluation (NDE) methods are used to audit building energy efficiency, such as fan pressurization (blower door test), ultrasound, and thermography [2–6]. In particular, drone-based infrared thermal (IRT) auditing is used to reconstruct 3D photogrammetric building energy models and to conduct energy simulations [7,8]. Researchers summarized different building energy loss types, such as unsealed windows, thermal bridges, moisture, etc. Thermal bridges' locations, boundary sharpness, generic structures, shapes, and types of building envelope materials are studied. In addition, researchers expect to use drone-based thermal cameras to detect thermal bridges from multiple buildings in a large district [8].

Drone-assisted energy audits are a novel approach to building energy audits that use unmanned aerial vehicles, also known as drones, to collect and analyze data on buildings' energy performance and efficiency. Drones offer multiple advantages over traditional data collection methods, including increased speed, enhanced safety, lower costs, and more comprehensive coverage of the building envelope and systems. Drones can access hard-to-reach areas, such as roofs, facades, windows, and vents, where most energy loss and waste occur. Drones can be equipped with various sensors and cameras, such as thermal, infrared, visible, multispectral, and hyperspectral sensors, to capture different types of data, such as temperature, humidity, pressure, airflow, solar radiation, and reflectance. Drones can also transmit the data in real time to a remote operator or a cloud server for further processing and analysis. Drone-assisted energy audits can provide more accurate and reliable data and more detailed and interactive visualizations than traditional data collection methods, which typically involve three main steps: planning, execution, and evaluation. Planning is the process of defining the objectives, scope, and level of the audit, as well as the drone specifications, flight parameters, and data requirements. Planning can be done using various software and tools, such as geographic information systems, building information modeling, and computer-aided design. Execution is the process of deploying and operating the drone to collect data on the energy performance and efficiency of the building. Execution can be done manually, semi-autonomously, or autonomously, depending on the level of human control and intervention. Evaluation is the process of analyzing and interpreting the data collected by the drone to evaluate the energy performance and efficiency of the building and identify energy problems and opportunities. Evaluation can be done using various software and tools, such as computer vision algorithms.

Despite their promise, drone-assisted inspections of buildings and infrastructure demand skilled pilots capable of multitasking. Operators need to maintain proficient flight control and remember observation sequences, adapt flight paths in real-time, and uphold safety measures [9–11]. Indeed, instructors and researchers continue to examine how "toolsets, skillsets, and mindsets" affect training outcomes in STEM education [12–17]. Second, drones are intended to partner with operators rather than simply complete operators' instructions [18]. The trust in HRI raises STEM educators' attention [19–22]. Third, Operators' decisions can be supported by AI, especially computer vision algorithms that process aerial images. For example, object detection [23] and segmentation [24] diagnose building roof heat loss [25], detect photovoltaic module defects [26], and discover wind turbine blade damage [27]. Fourth, drone-assisted energy audits and inspection requires experienced engineers. For example, skilled engineers know critical checkpoints, and they can reduce errors and reworks. Researchers have summarized the critical checkpoints of drone-assisted inspection for built environments and infrastructure systems.

In summary, drone-assisted energy audits have revolutionized assessing building envelope conditions. Nevertheless, the efficient and safe application of this technology depends not only on drone hardware and AI algorithms but also on the expertise and training of human operators, who must integrate systematic flight planning with knowledge of building physics and effective HRI. There is a lack of research about how human factors and personality affect drone-based inspection for building energy audits.

## 2.2 Human-Robot Interaction (HRI) for Inspection with Personality

The proliferation of robotic systems in industrial operations has necessitated advanced HRI frameworks capable of addressing the cognitive complexities inherent in infrastructure inspection workflows. Unlike structured manufacturing settings, civil infrastructure environments present dynamic spatial configurations, transient operational hazards, and heterogeneous defect morphologies - conditions that demand adaptive HRI architectures sensitive to human cognitive diversity.

### 2.2.1 Visual Inspection with HRI

Visual inspection is an object's cautious and critical assessment process regarding predefined standards [28,29]. Drury et al. [29] summarized five basic processes of inspection tasks – setting up, presenting, searching, making decisions, and responding - reveals critical pressure points in infrastructure diagnostics: (1) Temporal Constraints: Time-bound defect detection under resource limitations amplifies cognitive fatigue, particularly during sustained aerial inspections. Usually, the inspection tasks must be completed in a limited time and with restricted resources, and their complexity would be increased by the defects' various severity levels and multiple locations [30]. (2) Signal-to-Noise Challenges: Low-defect prevalence (typically <2% in building envelopes) coupled with high consequence costs necessitates optimized visual search strategies [15]. (3) Memory Load Dynamics: During the visual inspection, inspectors must concentrate, process, and transmit information exclusively using short-term (remembering which locations or components have been inspected) and long-term memory (recalling the initial conditions and standards) [31]. Extracting, understanding, and assessing inspectors' behavior and strategies can provide valuable insights for improving inspection performance and reducing errors.

Robots, vision, and sensing technologies have been used for visual inspection [32–34]. However, these systems will not merely replace human-based inspection because these systems cannot match human performances in tasks requiring flexibility and intelligence [35]. Emerging neuroeconomic studies demonstrate that individual differences in attentional control capacity and visual working memory span account for high variance in inspection accuracy across infrastructure domains [15]. These findings underscore the necessity for personality-aware HRI design in drone-assisted diagnostics, where adaptive human-machine collaboration can optimize workflow, minimize cognitive overload, and enhance defect detection capabilities. Developing intelligent HRI frameworks tailored to individual cognitive and perceptual tendencies will be instrumental in advancing drone-assisted infrastructure inspection.

### 2.2.2 The Role of Personality in HRI

Understanding the factors that impact HRI is crucial for designing effective and user-friendly robotic systems. Personality plays a significant role in HRI, influencing how individuals perceive and interact with robots. The MBTI is a widely used tool for assessing personality types [1]. The MBTI categorizes individuals based on four dichotomies: Extraversion (E) vs. Introversion (I), Sensing (S) vs. Intuition (N), Thinking (T) vs. Feeling (F), and Judging (J) vs. Perceiving (P). The permutation of these four dichotomies can form 16 distinct personality types, such as ISTJ, ENFP, and so on. These personality dimensions can significantly impact HRI in various ways.

Extraverts (E) are typically more outgoing and sociable, which may make them more comfortable and engaged when interacting with robots. They might prefer robots that offer interactive and dynamic experiences. Introverts (I), on the other hand, might favor robots that provide more structured and less stimulating interactions, valuing clarity and predictability. Sensors (S) prefer concrete information and practical applications, likely favoring robots that offer tangible and immediate benefits. They may respond well to robots designed for specific, practical tasks. Intuitive (N) are more comfortable with abstract concepts and future possibilities, possibly showing a preference for robots that embody innovative and experimental features. Thinkers (T) make decisions based on logic and objective analysis, often preferring robots that provide clear, rational responses and data-driven interactions.

Feelers (F) prioritize emotions and personal values in decision-making, potentially favouring robots that exhibit empathy and emotional intelligence. Judgers (J) prefer structure and organization, which may lead them to favor robots that offer predictable and orderly interactions. Perceivers (P) are more flexible and spontaneous, likely appreciating robots that can adapt to changing circumstances and offer varied experiences.

Understanding how different personality types interact with robots can help design more personalized and effective HRI systems. One of the challenges in this regard is to measure and model human personality in relation to robot characteristics and behaviours. Various methods have been proposed to assess personality types in HRI, such as questionnaires [40], behavioural observation [41], or physiological signals [42]. However, these methods are often subjective, intrusive, or time-consuming and may not capture a personality's dynamic and context-dependent nature. Moreover, most studies on personality and HRI focus on human-robot interaction in social or entertainment domains, such as education, health care, or gaming. There is a lack of research on how personality affects human-robot interaction in more task-oriented and professional domains, such as drone-based inspection tasks. There is limited literature on impact factor analysis and human personality modelling investigation for human-drone-environment for the built environment, infrastructure inspection, and related undergraduate education.

## 2.3 Simulation-based Learning Methods for Inspection Tasks

### 2.3.1 Immersive Simulation-based Learning with Behaviour Capturing

Traditional classroom training cannot repeat training modules for skill retention, build individual knowledge repositories [36], or provide real-time feedback [19]. Traditional training methods may also require trainees to travel [37]. Immersive simulation-based learning is an educational approach that uses computer simulations to create realistic scenarios for learners to practice their skills and knowledge in a safe and controlled environment. It can offer many benefits for inspection training, such as enhancing learners' motivation, engagement, feedback, and transfer of learning. It can also allow learners to experience various inspection tasks, defect types, environmental conditions, and autonomy levels without time, cost, and safety constraints. Moreover, it can facilitate the collection and analysis of learners' behavior data, such as eye movement, mouse clicks, keystrokes, and event logs, to evaluate their performance and strategies. Several studies have explored the use of immersive simulation-based training for inspection training in different domains, such as aviation, manufacturing, and medical fields. In the construction industry, researchers have developed a virtual environment and trained engineers to inspect highway construction [37], bridges [15], and nuclear power plants [38]. To enhance communications between humans and construction robots, researchers have proposed a cyberlearning platform [39], AI-assisted inspections [40], and Large Language Models, such as ChatGPT [41]. However, current studies on buildings and infrastructure inspection training with the support of simulation-based learning are inadequate.

Researchers have captured inspection behavior by tracking inspectors' eye movement [42,43], recording mouse inputs, and processing BIM event logs to investigate inspectors' strategies and decision-making processes [15] in simulations. Considering the inherent differences among inspectors, the significant discrepancy in inspectors' performance should be extracted and investigated [44,45]. Researchers have attempted to understand the experienced workers' inspection strategies using many process mining methods. For example, process mining has been employed across various domains by analyzing workers' event logs [46]. However, the quality control of the final results is challenged by the variability in experience and domain knowledge among workers and the significant individual subjective judgments involved [47,48].

Immersive simulation-based training could be a key means of transferring tacit knowledge. The actions and movements can be captured, thus allowing trainees to observe and experience the actions in high fidelity [49]. Additionally, the immersive learning modules could be used to train many individuals, lowering the training burden on senior inspectors. However, due to the complexity of inspection, simulation-based training for such tasks has not been investigated sufficiently to improve knowledge transfer efficiency and learning comfort. The drone-based inspection involves complex and dynamic interactions among humans, drones, and environments, which pose significant challenges for training and assessment. Therefore, there is a need to develop and evaluate simulation-based training methods that can effectively train and measure drone operators' inspection skills and strategies.

### 2.3.2 Participants' Performance Indicators and Measurement

Researchers have used physiological data (quantitative evaluation) and self-report measures or pre-defined indicators (qualitative evaluation) to measure immersive training performance. Physiological data, including neuropsychological data (electroencephalography (EEG) for the brain), electrocardiogram (ECG), eye tracking, skin temperature, and thoracic posture, have been investigated using wearable devices [65]. Particularly, heart rate variability (HRV) is a widely used physiological measure that reflects the autonomic nervous system's (ANS) regulation of the heart [50,51]. It is considered a reliable balance indicator between the sympathetic nervous system (SNS) and the parasympathetic nervous system (PNS). The SNS is responsible for the 'fight or flight' response, preparing the body for stressful situations by increasing heart rate, blood pressure, and energy availability. In contrast, the PNS is involved in 'rest and digest' activities, promoting relaxation, energy conservation, and digestion. PNS and SNS indexes are summarized in Table 1.

Table 1. Introduction of HRV Measurement Metrics

| HRV Measure | Description | Physiological Significance |
|---|---|---|
| Parasympathetic Nervous System Index | The PNS index measures parasympathetic cardiac activity, which influences HRV by decreasing heart rate, enhancing HRV via increased respiratory sinus arrhythmia (RSA), and reducing the ratio between lower and higher frequency oscillations in HRV time series [50,51]. | A PNS index value of zero indicates that the parameters reflecting parasympathetic activity are, on average, equivalent to those of the normal population [50]. Positive index values signify levels above the norm, and negative values indicate levels below. |
| Sympathetic Nervous System Index | The SNS index evaluates sympathetic cardiac activity, which typically increases heart rate, decreases HRV by reducing rapid RSA-related changes, and raises the ratio between lower and higher frequency oscillations in HRV data [50,51] | A SNS index of zero indicates average sympathetic activity compared to the norm. Positive values reflect sympathetic activity levels above the norm, while negative values indicate lower-than-average activity. |
| SD1, SD2 | SD1 measures the short-term variability in heart rate and reflects the activity of the PNS. SD2 measures the long-term variability in heart rate and is influenced by both SNS and PNS activity. | SD1 is strongly influenced by parasympathetic activity, and SD2 reflects overall variability |
| Baevsky's Stress Index | An index used to quantify interbeat interval shapes and distribution during sympathetic activation and is based on a statistical analysis of histogram of the intervals between successive heartbeats (also called RR-Intervals) distribution | Reflects stress-related sympathetic activation levels |

Understanding these HRV metrics and their implications can aid in the early detection and management of stress-related conditions. Therefore, it is necessary to investigate the validity and reliability of HRV as a measure of cognitive workload in immersive simulation-based training for building energy audits and to compare it with other methods, such as subjective ratings and performance indicators. There is a lack of objective and reliable methods to measure cognitive workload in immersive simulation-based training for building energy audits, a key factor affecting trainees' learning outcomes and safety performance.

## 2.4 Research Gaps

Based on the literature review, we identify the following research gaps that motivate our study:

(1) *Limited Multi-dimensional Performance Metrics for VR-Based Drone Training:* While immersive technologies have been applied to various training scenarios, there is a significant gap in establishing comprehensive metrics that integrate behavioral, physiological, and operational dimensions for evaluating performance in drone-assisted building inspection training. Current evaluation frameworks typically focus on single performance measures rather than holistic assessment approaches that capture the complex interplay between flight control proficiency, cognitive processing, and physiological responses.

(2) *Insufficient Comparative Analysis of MBTI Influence on Training Performance:* The role of personality traits, as classified by the MBTI, in immersive simulation-based training for building energy audits, remains underexplored. There is a lack of comparative studies validating how different MBTI personality types of influence trainees' performance, learning experiences, and decision-making processes in drone-assisted inspections.

(3) *Lack of Personalization Frameworks for Technical VR Training*: Current approaches to VR-based technical training often employ one-size-fits-all methodologies that fail to account for individual differences in cognitive processing, stress management, and operational preferences. There is a pressing need for research that establishes foundations for personalizing training experiences based on learners' unique psychological and physiological profiles to optimize skill acquisition and knowledge transfer in specialized technical fields such as building energy auditing

## 3. Methodology

### 3.1 The Framework of the Drone Building Heat Loss Test Case

We designed a framework for drone control training in a VR environment to understand the interaction between humans and drones for building heat loss detection. Fig. 1 shows the proposed framework, including (1) VR Environments Setup, (2) Preparation, (3) Pre-Surveys, (4) Experiments, (5) Post Survey, and (6) Data Analysis. Using this framework, we built a test case for a drone-based building energy audit using the immersive simulation for training purposes. The test case involves using a virtual drone with thermal cameras and monitors to inspect the building's thermal performance and identify potential energy losses and defects. The drone can be controlled remotely by an operator in a virtual environment. The participants in this study are novice drone operators who need to learn how to perform a drone-assisted building energy audit. The immersive simulation provides a realistic and safe environment for the participants to practice their skills and gain feedback on their performance.

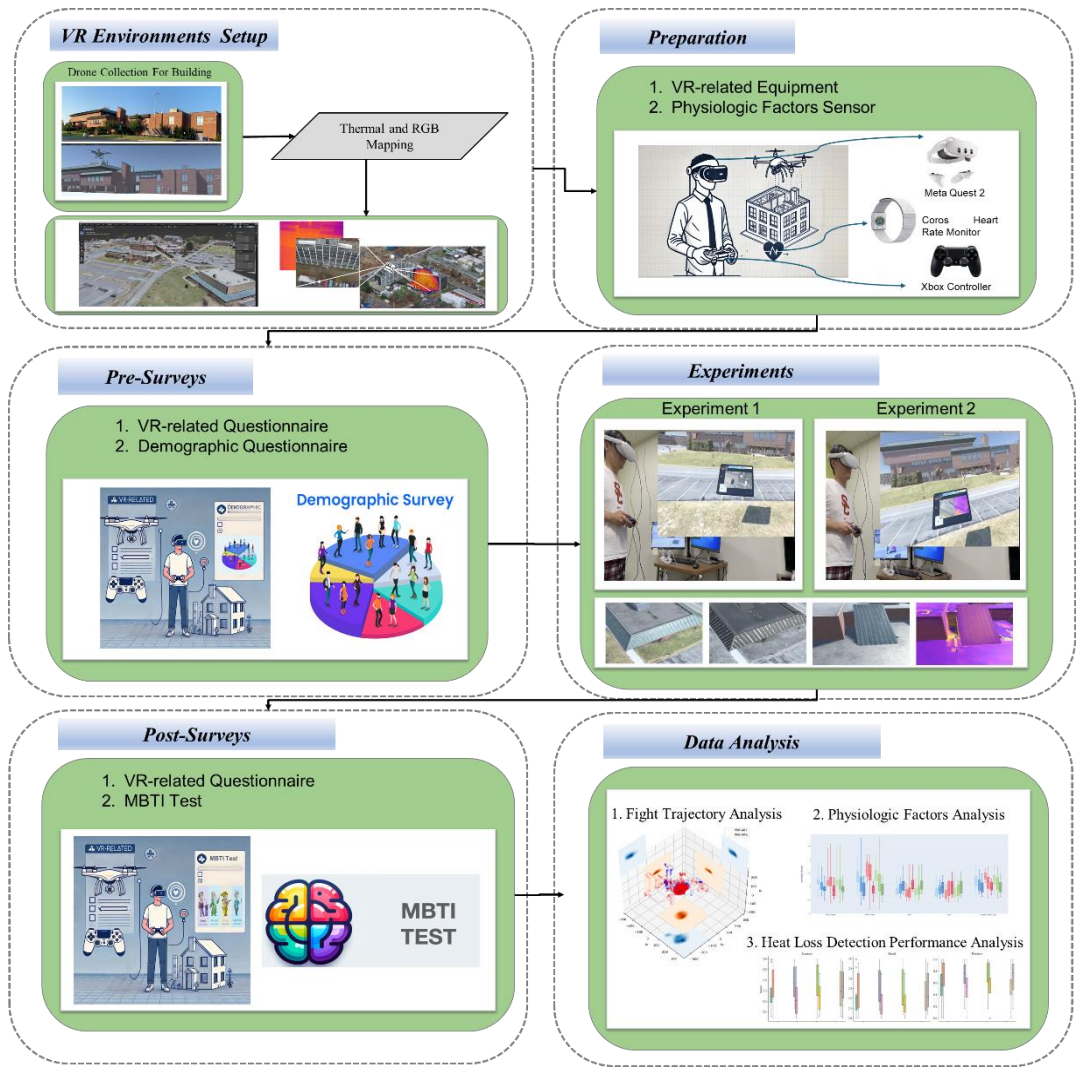

Figure 1. Framework for the proposed VR-based workforce training on human–drone interaction

In the (1) VR Environments Setup, we used a real drone with RGB and thermal cameras to capture onsite RGB and thermal images of a Western New England University (WNE) building. Then we digitalized the building in the VR environment using photogrammetry technologies. This procedure ensured that the heat loss information was based on a real case rather than synthetic data.

In the (2) Preparation stage, participants are equipped with a VR headset (*Meta Quest* 3 in this study) and game controllers (*PlayStation 4* or Xbox controller). This stage involves setting up the VR equipment, ensuring the participants sign an informed consent approved by the WNE's Institutional Review Board (IRB), and being ready for the experiments.

Following this, in the (3) Pre-Surveys stage, the participant completes VR experiment-related and demographic questionnaires to gather baseline data. It includes participants' educational background, demographic questionnaires, and past experience with VR to gather baseline data. The questionnaires can be found in the attachments.

The (4) Experiments stage of the workflow involves two experiments (Exp1: initial free flight and Exp2: heat loss detection), during which participants interact with a VR environment to control a drone to detect heat loss in the built environment. we leveraged the Unity Engine to develop and test the learning environment. In Exp1, participants familiarize themselves with the drone control setting and are allowed to pilot the drone at leisure to explore the WNE buildings and campus in the virtual environment. Meanwhile, their physiological data is gathered for baseline purposes. In Exp2, participants are multitasked by maneuvering the drone around a building to perform a thermal inspection for heat loss detection. Therefore, participants must balance safe drone operation and accurate building inspections by examining thermal images through a monitor.

After the experiments, the (5) post-survey stage involves administering another VR-related questionnaire to report their comfort level, the perceived difficulty of the tasks, any technical issues encountered, and their overall experience with the VR simulation. The post-survey also includes a MBTI test to gather personality data. The questionnaires can be found in the attachments. Participants fill out a survey while in a relaxed state, reflecting on their experience.

The final stage, (6) Data Analysis stage, involved processing and interpreting flight trajectories, physiological measurements (e.g., heart rate variability), and participants' performance in detecting heat loss. A Coros heart rate monitor continuously tracked physiological responses throughout the experiments, enabling a detailed understanding of participants' stress levels and cognitive load, especially in the (4) experiments stage and (5) post-survey stage.

## 3.2 Experiment Establishment

### 3.2.1 Software - VR Environment Establishment

In the VR environment establishment, we used a real drone, *DJI Mavic* 3, with RGB and thermal cameras to capture onsite overlapped RGB and thermal images of a WNE building. These overlapped images were utilized to reconstruct 3D mesh models of the WNE campus by using photogrammetry technologies. After 3D RGB mesh models were reconstructed, we projected temperature information from thermal images onto the mesh models to generate thermographic models. Thus, models for virtual environments with RGB and thermal information were established. Related technologies were elaborated in our previous work [6,47,48], as shown in Figure 2 (a). Figure 2 (b) shows the participants' views in the virtual environment. There was a virtual iPad monitor showing the views of the virtual drone's RGB and thermal camera. While participants were operating the drone, the iPad monitor showed the drone camera's view simultaneously. For the thermal camera, darker purple colors represent lower temperatures, while brighter yellow colors represent higher temperatures. Because the thermographic data originate from real drone flights, heat loss areas in the simulation accurately reflect the building's actual thermal behavior. This authenticity allows participants to experience realistic conditions for training in energy audits.

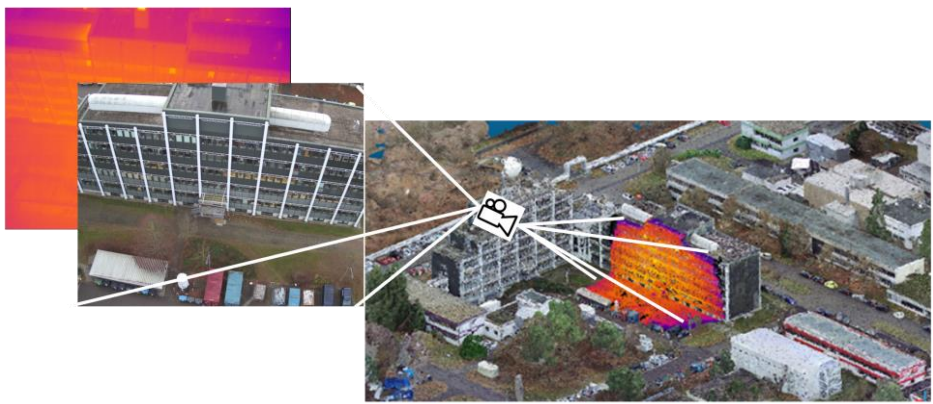

(a)  RGB and Thermal Mapping by Using Photogrammetry Technologies

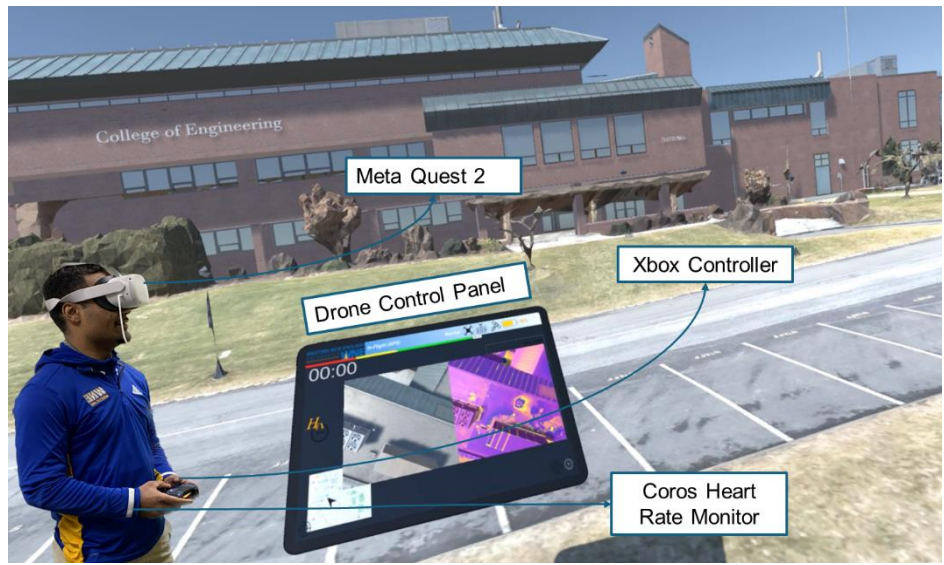

(b) Visualization of the Drone Control Pannel in Unity Game Engie

Figure 2. VR Environment Establishment for Heat Loss Detection

### 3.2.2 Hardware - Equipment Setup

As shown in Figure 2 (b), in the Preparation stage, the experiment system setup consists of a *Meta Quest* 3 head-mounted display (HMD) with integrated headphones and microphone, a *PlayStation 4* or Xbox controller, and a *Coros* Heart Rate Monitor. The HMD features a resolution of 1440 x 1600 pixels per eye, a refresh rate of 90 Hz, and a 110-degree field of view, providing an immersive visual experience. The integrated audio system delivers spatial sound, enhancing the sense of presence in the virtual environment. The controller is equipped with buttons and triggers, simulating the controlling operations of the drone and allowing precise interaction with the virtual environment. The *Coros* Heart Rate Monitor is used to track the participants' physiological responses during the simulation. It measures HRV, providing insights into the participants' stress levels and engagement with the tasks, which is crucial for assessing the participants' performance. The HMD and other hardware are connected to a high-performance computer with specifications including an *Intel* Core i9 processor, 32GB of RAM, and an *NVIDIA* RTX 3080 GPU. This setup ensures smooth simulation rendering and quick processing of user inputs.

The simulation software is developed using *Unity* 3D, a game engine that supports VR development and integration. The simulation scenario is based on a real-world task of inspecting a building for heat loss using a thermal camera. Participants can move around the building and use the controller to point and shoot the thermal images from different views of building envelopes. The simulation provides visual and auditory feedback to participants, such as the thermal images of the building envelopes, the temperature reading, and the sound of the camera shutter. The simulation aims to identify areas of heat loss in a virtual building using a thermal camera. The system records participants' actions and provides real-time feedback to improve their performance. The data collected from the heart rate monitor, along with the performance metrics, is analyzed to evaluate the effectiveness of participants' performance.

### 3.2.3 Participants

We recruited 26 Western New England University participants to participate in the experiment. The participants had various backgrounds with and without Science, Technology, Engineering, and Mathematics (STEM). The participants' ages ranged from 18 to 35 years old, with an average of 23.6 years. All participants had normal or corrected-to-normal vision and no history of neurological disorders. Participants must report the climate zones in their past and current living environments.

## 3.3 Experiment Settings

As introduced in the experimental stage of the framework, two experiments (Exp1: initial free flight and Exp2: heat loss detection) were used in this study. Each was designed with varying tasks and complexity to thoroughly assess the participants' ability to perform drone-based building energy audits.

In Exp1, participants familiarize themselves with drone operations in a virtual environment. Participants follow a standard procedure for a drone-based building energy audit within a specified time limit. This procedure begins with preparing the drone, including checking the battery level and verifying camera settings. They are allowed to fly the drone freely on the WNE campus. They can practice basic maneuvers, learn to control the drone's speed and direction, and become familiar with the VR interface and controls. This stage occurs in an open virtual area with minimal obstacles to prevent overwhelming the participants, allowing them to build confidence and proficiency in handling the drone.

In Exp2, participants conduct a simulated heat loss energy audit of a building using a drone. During this stage, participants must navigate the drone around a selected virtual building to capture thermal images of critical areas from building envelopes. They can define drone inspection strategies and waypoints to ensure thorough coverage of the building's exterior. Exp2 introduces intermediate challenges, such as moderate wind or narrow passages, to simulate real-world conditions, testing the participants' ability to maintain control and accurately perform the survey. Upon completion of the drone flight, participants retrieve the collected thermal data. They analyze the thermal images to identify heat loss from the images taken and compile their findings into a comprehensive energy audit report.

## 3.4 Data Analysis

The data analysis section comprises three main areas: (1) fight trajectory analysis, (2) HRV assessment, and (3) heat loss detection performance analysis. Additionally, results from the MBTI assessment are integrated into these three areas to explore the influence of personality traits on participants' performance and behavior.

### 3.4.1 Fight Trajectory Analysis

The flight trajectory analysis examines how participants pilot the virtual drone within the simulation. We track each drone's movement over time by collecting spatial coordinates and temporal data through the VR system. This enables us to: (1) Identify Inspection Strategies: Determine how participants navigate the drone to locate and inspect key areas of interest. (2) Assess Task Efficiency: Evaluate participants' paths, highlighting potential inefficiencies such as excessive backtracking or repetitive scanning. (3) Visualize Movement Patterns: Generate visual representations of flight paths to identify commonly used routes and recurring strategies. Furthermore, MBTI classifications are incorporated to discern whether personality traits correlate with specific piloting behaviors. For instance, individuals with high extraversion may exhibit broader exploration patterns, whereas introverts might adopt more methodical routes. This analysis offers targeted insights for refining future training, optimizing inspection strategies, and tailoring instruction based on trainees' personality profiles.

$$Spatial\ Dispersion: \sigma_a^2 = \frac{1}{N}\sum_{i=1}^{N}(ai - \mu_a)^2 (a \in \{x, y, z\}) \qquad (1)$$

This metric calculates the variance along each axis (x, y, z), representing how far participants move from the mean position .

$$Total\ Variance: \sigma_{total}^2 = \frac{1}{3}(\sigma_x^2 + \sigma_y^2 + \sigma_z^2) \qquad (2)$$

We derive a single value that indicates overall dispersion in 3D space by averaging the variance across all three axes. A higher total variance suggests more wide-ranging or scattered flight paths.

$$Global\ Coordination\ Index: \Gamma = \frac{1}{d*d}\sum_{i=1}^{d}\sum_{j=1}^{d} Cov(a_i, a_j)\ (d=3) \qquad (3)$$

The global coordination index measures how strongly movements along different axes covary. A higher value indicates coordinated motion across axes (e.g., diagonal or curved flight) rather than purely axis-aligned maneuvers.

$$Global\ Abs\ Coordination\ Index: \Gamma_{abs} = \frac{1}{d*d}\sum_{i=1}^{d}\sum_{j=1}^{d}|Cov(a_i, a_j)|\ (d=3) \quad (4)$$

This variant uses the absolute value of covariance, capturing total coordination magnitude regardless of positive or negative correlations. High values reflect complex flight maneuvers, where participants coordinate multiple axes in tandem.

### 3.4.2 HRV Assessment in Different Immersive Simulation Scenarios

HRV assessment provides valuable insights into the physiological responses of participants under different immersive simulation scenarios. We can measure the participants' stress levels by analyzing HRV data collected using the Coros Heart Rate Monitor. Key HRV metrics include the PNS Index, SNS Index, SD1, SD2, and SI. The theories behind these metrics are summarized in the literature review section. The analysis involves comparing these HRV metrics across different scenarios and differentiating the metrics based on participants' MBTI personality classifications. This approach helps determine how factors such as task complexity and personality traits affect the inspection performances, which allows us to identify optimal conditions for training and improve the design of simulation scenarios.

### 3.4.3 Heat Loss Detection Performance Analysis

Inspectors aim to find the defects among all the elements. Compared to the normal elements (non-heat loss areas), the number of abnormal elements is rare. Heat loss detection performance analysis focuses on evaluating the effectiveness and accuracy of trainees in identifying areas of heat loss in the virtual environment. Therefore, the commonly used metrics in anomaly detection are also suitable for evaluating the inspectors' performances. In addition, as shown in Figure 4, by comparing the inspectors' final inspection of the heat loss and the ground truth of heat loss, inspectors' performances could be evaluated. Key parameters analyzed include the number of shots (the total number of thermal images taken by the trainee) and accuracy of shots (the proportion of correctly identified areas of heat loss compared to predefined correct locations). Additionally, we categorize these performance metrics based on the trainees' MBTI personality classification to explore how personality traits influence performance in heat loss detection. This analysis helps assess the trainees' proficiency in using the thermal camera and interpreting thermal images. It identifies areas where additional training or adjustments to the simulation scenario may be needed to improve performance.

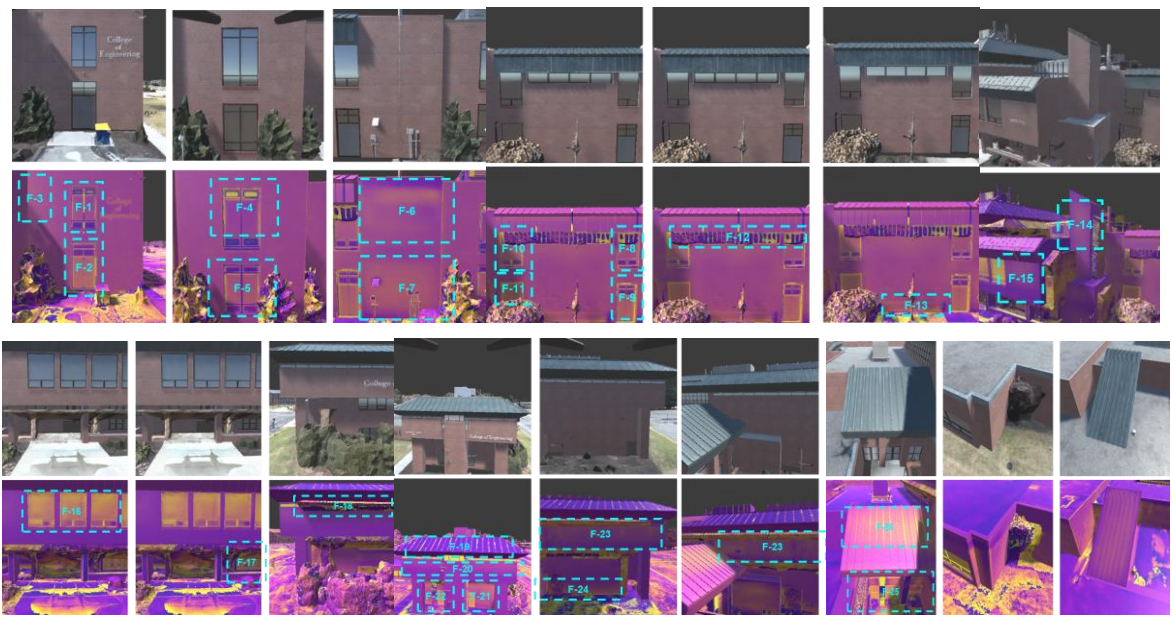

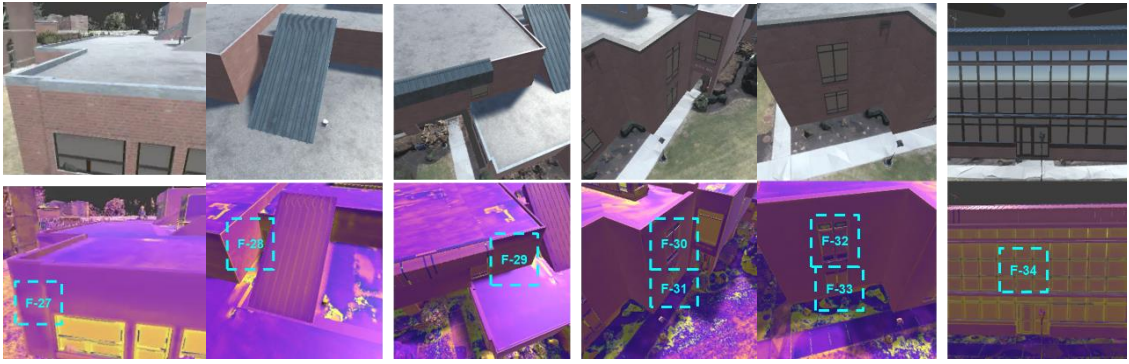

(a) Façade Heat Loss Ground Truths

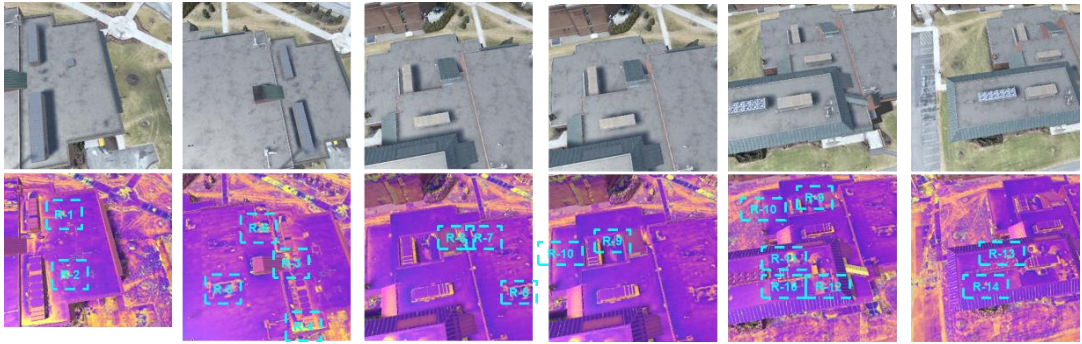

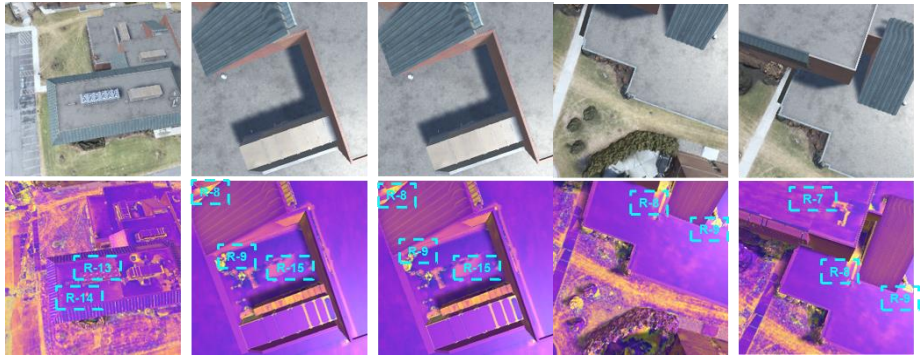

(b) Roof Heat Loss Ground Truths

Figure 4. Ground Truth of Heat Loss in VR Environments

In this research, heat loss is known as a positive class, whereas the heat-loss-free elements belong to a negative class. To comprehensively evaluate the performance of heat loss detection in drone-assisted inspections, we further define the following metrics based on the confusion matrix:

$$Accuracy = \frac{True\ Positive + True\ Negative}{Total\ Elements} \quad (5)$$

$$Recall = \frac{True\ Positive}{(True\ Positive + False\ Negative)} \quad (6)$$

$$Precision = \frac{True\ Positive}{(True\ Positive + False\ Positive)} \quad (7)$$

$$F1 = \frac{2 * Recall * Precision}{(Precision + Recall)} \quad (8)$$

To further evaluate the efficiency and practicality of drone-based heat loss inspections, we introduce two additional metrics and a detailed error classification scheme:

Coverage Rate (CR) is defined as the ratio of detected heat loss instances to the total number of images captured; this metric quantifies the inspector's ability to prioritize meaningful data collection. A higher

CR indicates efficient drone operation that focuses on areas with potential defects, minimizing redundant or irrelevant captures.

$$Coverage\ Rate\ (CR) = \frac{Number\ of\ detected\ heat\ loss\ instances\ (True\ Positives)}{Total\ number\ of\ images\ captured\ (N)} \quad (9)$$

The false rate of wrongly taken images (FAR) measures the proportion of images containing incorrectly flagged heat loss areas (false positives) relative to the total captured images. A low FAR reduces unnecessary post-processing efforts and resource waste.

$$False\ Rate\ (FAR) = \frac{Number\ of\ Misclassified\ Images\ (False\ Positives)}{Total\ number\ of\ images\ captured\ (N)} \quad (10)$$

In addition to these quantitative metrics, Table 2 outlines a classification scheme for analyzing misclassifications based on image content. This classification of errors provides insights into participants' specific types of mistakes, allowing for targeted improvements in VR-based training. It helps differentiate between errors caused by misinterpretation of thermal images versus operational errors in drone navigation. By integrating both standard classification metrics and additional performance indicators, this study offers a comprehensive assessment of heat loss detection accuracy and error patterns to further investigate how personality traits might correlate with different error types and overall performance.

**Table 2. Classification of Error Types**

| Error Type | Definition | Example |
| --- | --- | --- |
| **Treat AC as heat loss** | Incorrectly classifying air conditioning units as heat loss. | A participant mistakenly marks an HVAC unit as a heat leak. |
| **Wrong Building** | Capturing and analyzing the wrong building. | The drone operator inspects a neighboring building instead. |
| **No Info (Regular Roof/Wall)** | Capturing areas without any heat loss indications. | A participant captures a well-insulated roof with no defects. |
| **Sky** | Taking images of the sky instead of building surfaces. | The drone camera points upwards, capturing only the sky. |
| **Floor** | Capturing the ground instead of heat loss-prone areas. | The operator accidentally takes images of pavement or grass. |

## 4. Experimental Results and Discussions

This section synthesizes the statistical evidence linking MBTI personality traits to human-drone interaction patterns through three analytical lenses: (1) spatial navigation dynamics derived from 3D flight trajectory mining, (2) psychophysiological adaptation captured via autonomic nervous system responses, and (3) task performance metrics quantifying inspection accuracy and error profiles. Our tripartite analysis reveals how cognitive-behavioral dispositions systematically modulate operational strategies, stress resilience, and diagnostic efficacy in VR-based thermal inspection training. The findings are contextualized within human factors engineering and adaptive learning frameworks, offering actionable guidance for designing personality-aware training protocols.

### 4.1 Flight Trajectory Analysis

We monitored participants' flight paths and depicted the spatial patterns of different MBTI personality types along the four dichotomies in Figures 5 and 6 and Table 3. The literature indicates that the four dichotomies are Extraversion (E) versus Introversion (I), Sensing (S) versus Intuition (N), Thinking (T) versus Feeling (F), and Judging (J) versus Perceiving (P). In evaluating participant performance, we strictly classified participants into each corresponding dichotomy. Each graph in Figure 5 clearly distinguishes between two contrasting personality types in the Exp 1.

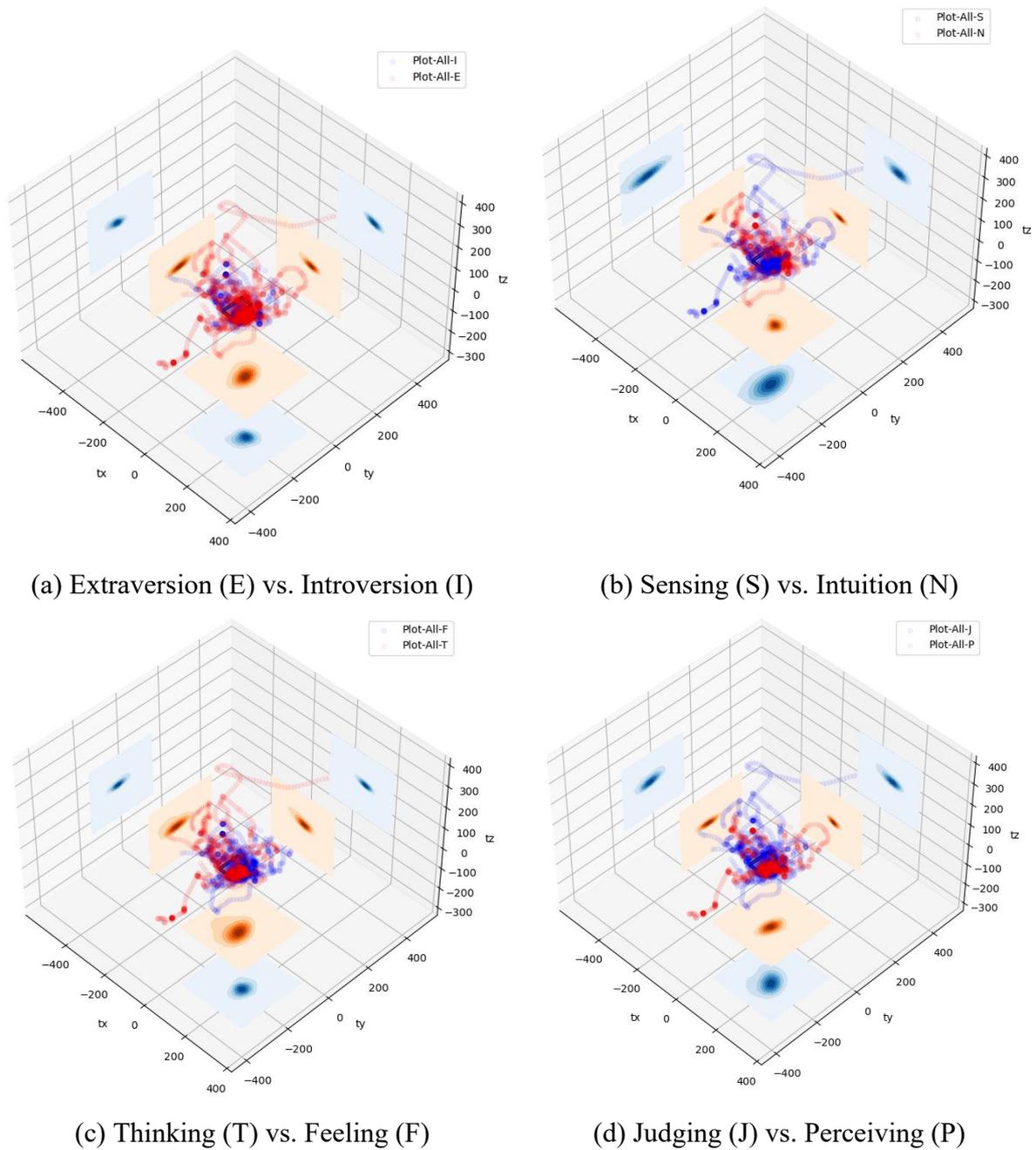

(a) Extraversion (E) vs. Introversion (I)      (b) Sensing (S) vs. Intuition (N)

(c) Thinking (T) vs. Feeling (F)      (d) Judging (J) vs. Perceiving (P)

Figure 5. Flight trajectory analysis for the exploration stages (Exp 1)

**Table 3. Comparison of Flight Trajectory with MBTI**

| MBTI | Dimension | Experiment | Variance x | Variance y | Variance z | Variance mean | Avg Cov (Global Coordination Index) | Avg Abs Cov (Global Abs Coordination Index) |
|---|---|---|---|---|---|---|---|---|
| E/I | E | Exp 1 | 514 | 11567 | 1457 | 6179 | 1838 | 2893 |
|  |  | Exp 2 | 3054 | 4807 | 902 | 2921 | 930 | 1491 |
|  | I | Exp 1 | 2833 | 1728 | 952 | 1838 | 363 | 878 |
|     |   | Exp 2 | 2349 | 2823  | 2448 | 2540 | 667  | 1474 |
|     | N | Exp 1 | 3779 | 2152  | 1041 | 2324 | 422  | 1128 |
| N/S |   | Exp 2 | 1976 | 2288  | 932  | 1732 | 659  | 853  |
|     | S | Exp 1 | 5764 | 17836 | 1572 | 8391 | 2784 | 3915 |
|     |   | Exp 2 | 3832 | 6404  | 2158 | 4131 | 1061 | 2228 |
|     | F | Exp 1 | 2446 | 2560  | 634  | 1880 | 612  | 818  |
| F/T |   | Exp 2 | 2221 | 2310  | 1019 | 1850 | 652  | 859  |
|     | T | Exp 1 | 6838 | 15383 | 1921 | 8047 | 2187 | 3891 |
|     |   | Exp 2 | 3477 | 5700  | 2011 | 3729 | 1023 | 2114 |
|     | J | Exp 1 | 5577 | 6744  | 1396 | 4572 | 1259 | 2386 |
| P/J |   | Exp 2 | 2633 | 2655  | 1676 | 2322 | 744  | 1391 |
|     | P | Exp 1 | 2641 | 9006  | 1047 | 4232 | 1279 | 1690 |
|     |   | Exp 2 | 3031 | 6276  | 1249 | 3519 | 955  | 1657 |

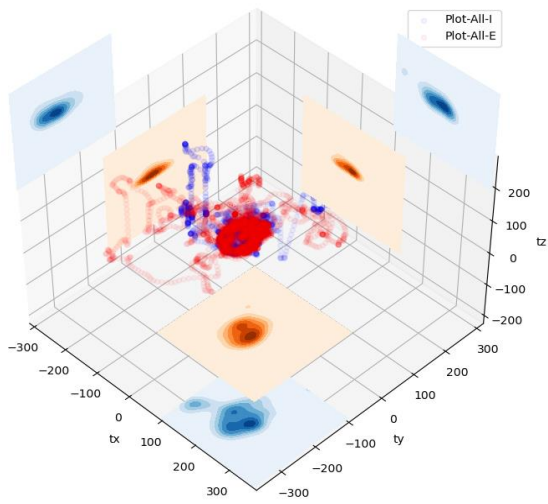

(a) Extraversion (E) vs. Introversion (I)

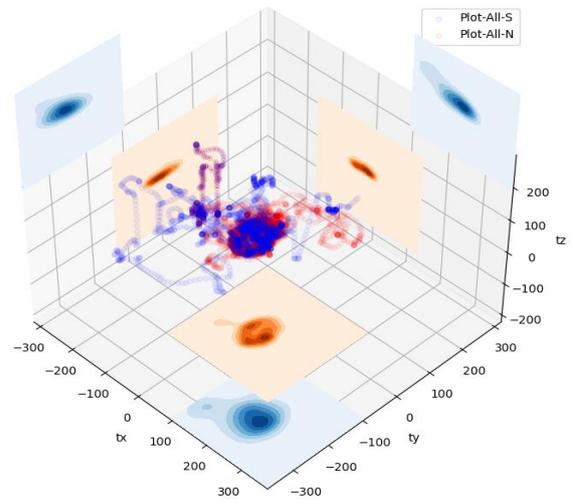

(b) Sensing (S) vs. Intuition (N)

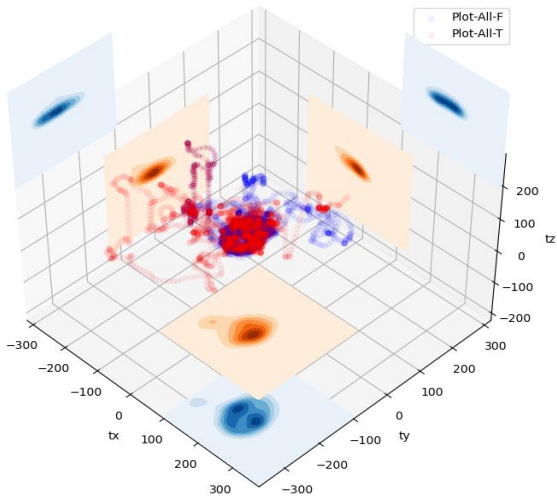

(c) Thinking (T) vs. Feeling (F)

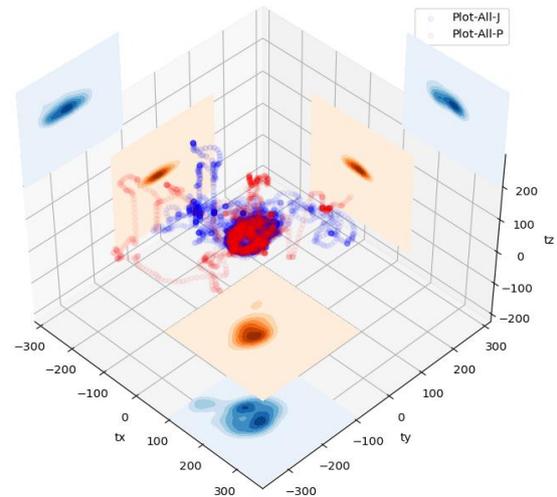

(d) Judging (J) vs. Perceiving (P)

Figure 6. Flight trajectory analysis for the heat loss detection stage (Exp 2)

### 4.1.1 Extraversion (E) vs. Introversion (I)

In the *Extraversion* (E) vs. *Introversion* (I) plot, in exploration stage (Figure 5.a) red points represent Extraversion while blue points represent Introversion, revealing unique movement patterns in the three-dimensional space. *Extraversion* trajectories are more scattered and cover a wider area, indicating a tendency towards outward exploration, whereas *Introversion* trajectories are more concentrated, suggesting inward focus and depth. This observation was also reflected in Table 3, as the Extraversion group has an absolute average variance of 2892.68, which was higher than the Introversion group's value of 878.25.

In heat loss detection Exp2 (Figure 6.a), *Extraversion* maintained their broader exploration patterns but showed more focused scanning behavior, covering more ground but with a purpose-driven approach (with an absolute average variance of 1490.62). Introversion, who initially had more concentrated and focused paths, continued to exhibit detailed and methodical approaches, often revisiting specific areas to ensure a thorough inspection (with an absolute average variance of 1474.48).

### 4.1.2 Sensing (S) vs. Intuition (N)

Similarly, the Sensing (S) vs. Intuition (N) plot differentiates between Sensing and Intuition with red and blue points, respectively, highlighting their divergent information processing approaches. In the exploration stage (Figure 5.b), Sensing trajectories showed remarkably higher variance across all dimensions, with a variance mean (8390.68) nearly four times that of Intuition types (2323.91). This is particularly evident in the y-dimension, where Sensing variance (17836.28) far exceeded Intuition variance (2151.56). Coverage metrics reinforced this pattern, with Sensing types demonstrating substantially higher average coverage (2783.97 vs. 422.25) and absolute average coverage (3915.16 vs. 1127.59).

In heat loss detection Exp 2 (Figure 6.b), this pattern persisted, with Sensing continuing to favor broader exploration with higher variance mean (4131.29 vs. 1731.92) and significantly higher coverage metrics (absolute average coverage: 2228.25 vs. 853.30). This suggests that Sensing types, contrary to their theoretical preference for concrete information, demonstrated more comprehensive spatial exploration, possibly reflecting their focus on gathering complete sensory data from the environment rather than relying on abstract patterns or intuitions about where defects might occur.

### 4.1.3 Thinking (T) vs. Feeling (F)

In Figure 5 (c), the Thinking (T) vs. Feeling (F) plot contrasts the decision-making styles and emotional responses of these types, again using red for Thinking and blue for Feeling. Thinking trajectories show substantially higher variance across all dimensions, with a variance mean (8047.39) over four times that of Feeling types (1880.15). This is particularly pronounced in the y-dimension (15383.38 vs. 2559.84) and reflects in their coverage metrics, where Thinking types demonstrate much higher absolute average coverage (3891.11 vs. 817.88). This suggests Thinking types may adopt more comprehensive scanning strategies to gather complete data before analysis.

During the heat loss detection task in Figure 6 (c), Thinking types maintained their broader exploration patterns with a higher variance mean (3729.22 vs. 1850.18) and more than double the absolute average coverage (2113.86 vs. 859.46). This suggests that Thinking types employed systematic but extensive coverage strategies, methodically examining the building's surfaces across a wider area, while Feeling types adopted more focused and potentially intuition-driven approaches to identify areas of concern.

### 4.1.4 Judging (J) vs. Perceiving (P)

Finally, in Figure 5 (d), the Judging (J) vs. Perceiving (P) plot demonstrates the structural versus spontaneous approaches of these types with red and blue points. Unlike other dimensions, the variance differences between Judging and Perceiving were less pronounced, with Judging showing slightly higher variance mean (4572.22 vs. 4231.57). However, Judging showed notably higher x-dimension variance (5576.91 vs. 2641.15), while Perceiving showed higher y-dimension variance (9006.46 vs. 6743.53). Coverage metrics were remarkably similar (absolute average coverage: 2386.05 vs. 1690.35), suggesting both types achieved comparable spatial exploration despite different trajectory patterns.

In Figure 6 (d), during the heat loss detection task, Perceiving types showed higher variance mean (3518.51 vs. 2321.50) and significantly higher y-dimension variance (6275.85 vs. 2655.24), indicating more lateral movement. However, Judging types maintained higher z-dimension variance (1676.45 vs. 1248.80), suggesting more methodical vertical exploration. Coverage metrics remained fairly similar (absolute average coverage: 1391.39 vs. 1657.11), indicating that both types achieved comparable inspection coverage through different spatial movement strategies.

### 4.1.5 Integrated Observations

In summary, if participants, they tend to participate in a wider area and engage in outward exploration. If they are Introversion, they tend to focus inwardly and explore in depth within a more concentrated area. If participants have a Sensing preference, they likely follow more linear and clustered trajectories, favoring concrete and practical information. In contrast, those with an Intuition preference have more dispersed trajectories, reflecting abstract and innovative thinking. Individuals with a Thinking preference exhibit more structured and systematic paths, showing a logical decision-making style, whereas those with a Feeling preference have more fluid and varied paths, indicating a values-driven and empathetic approach. Finally, those with a Judging preference demonstrate predictable and orderly trajectories, reflecting their preference for planning and organization, while individuals with a Perceiving preference show flexible and adaptable trajectories, indicative of their spontaneous approach. Task-specific demands (e.g., heat loss detection in Exp2) generally reduced variance differences between personality dimensions compared to free exploration (Exp1) but maintained consistent relative patterns. These findings highlight how personality traits fundamentally shape flight strategies and spatial exploration patterns, with implications for training in drone-assisted building audits. Understanding these inherent tendencies can inform customized training interventions that leverage individual cognitive-behavioral profiles for both general exploration and targeted detection tasks.

### 4.2 Physiologic Factors Analysis

Figure 7 illustrates a comparative analysis of physiological responses during different stages of drone operation, framed within the MBTI categorizations. These stages include Exp 1 (exploration), where participants freely operate drones to become accustomed to the controls; Exp 2, comprising tasks like navigating drones around buildings for heat loss detection; and the post-experiment relaxation period. The box plots depict normalized shifts in five physiological markers: PNS Index, SNS Index, SD1, SD2, and SI. Each segment is differentiated by color, signifying the distinction between four dimensions of MBTI profiles and aggregate data encompassing all types.

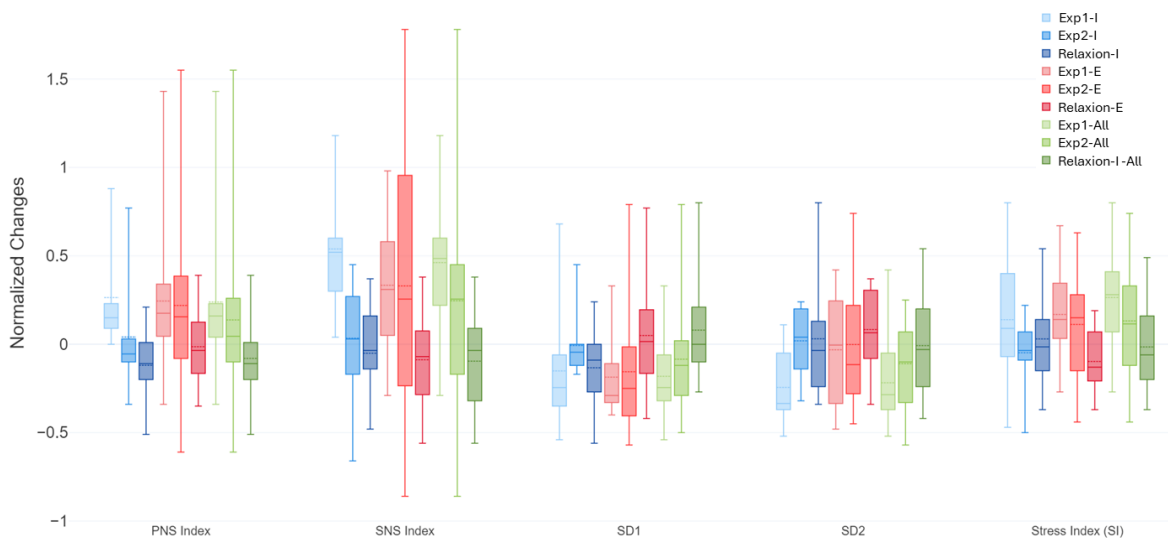

(a) Comparison of the physiologic factors for the Extraversion (E) and Introversion (I)

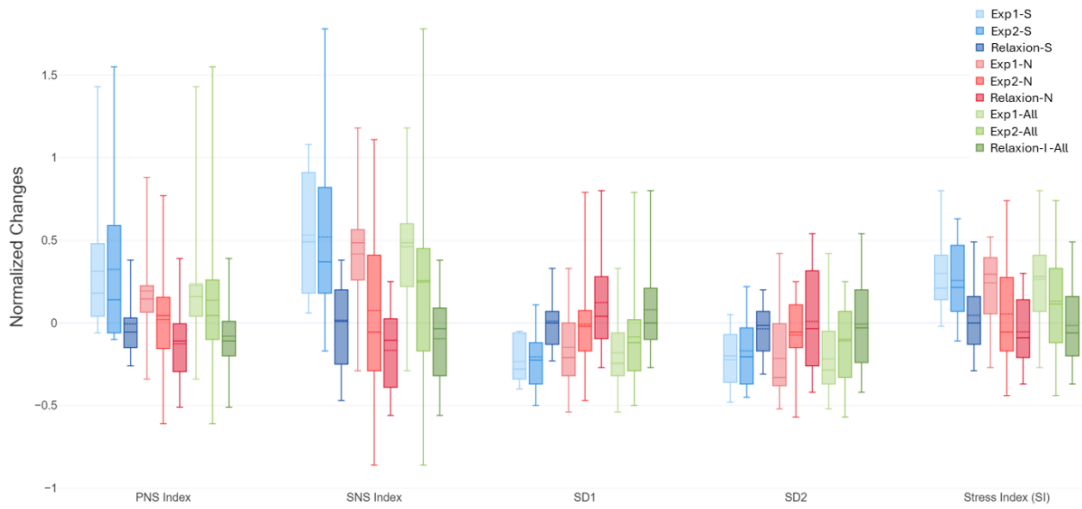

(b) Comparison of the physiologic factors for the Sensing (S) vs. Intuition (N)

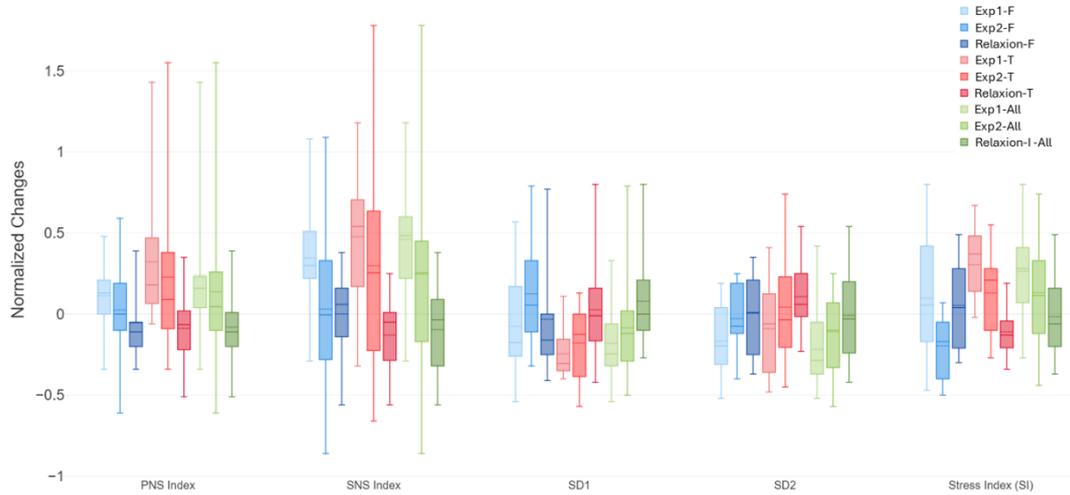

(c) Comparison of the physiologic factors for the Thinking (T) vs. Feeling (F)

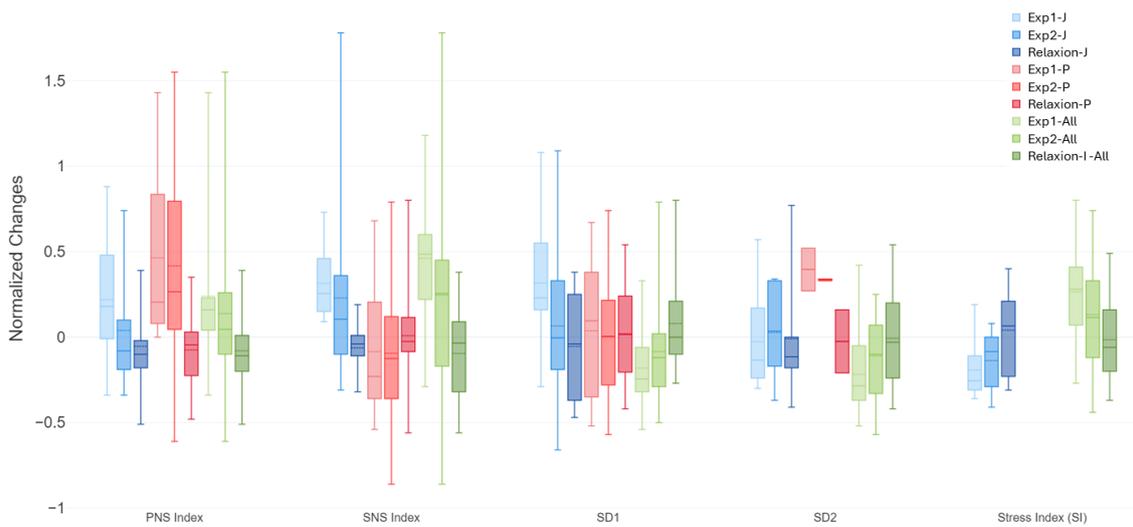

(d) Comparison of the physiologic factors for the Judging (J) and Perceiving (P)

Figure 7. Comparison of the physiologic factors for three stages (Exp 1, Exp 2 and relaxation) with MBTI

### 4.2.1 Extraversion (E) vs. Introversion (I)

Regarding Extraversion (E) and Introversion (I) typologies in Figure 7 (a), during Exp1 (initial free flight), varying physiological reactions are seen, chiefly in PNS and SNS indices. Extraverts (in red and pink) differ from Introverts (in light blue and blue) — showing wider normalized change ranges, implying more intense interaction and arousal during free-flying, PNS Index mean of 0.24 (std 0.39), and an SNS Index mean of 0.39 (std 0.35).

Exp 2 (heat loss detection) brings about heightened responses in all markers due to the demanding nature of the heat loss detection activity. Extraverts (E) maintain moderate PNS (mean 0.21, std 0.50) and SNS (mean 0.39, std 0.70) activity, along with a Stress Index near 0.19 (std 0.36). Introverts (I) show lower PNS activity (mean 0.04, std 0.29) and less SNS activation (mean 0.05, std 0.36), yet they display overall stress levels of 0.05 (std 0.21). These metrics suggest Extraverts may cope with higher variability during the task, whereas Introverts maintain more subdued but steady responses.

In the post-survey relaxation phase, both groups trend back toward negative or near-zero PNS and SNS Index values, indicating reduced stress. However, Extraverts show SI at –0.05 (std 0.23), compared to Introverts at 0.03 (std 0.23), implying slightly different relaxation trajectories.

### 4.2.2 Sensing (S) vs. Intuition (N)

Comparing the Sensing (S) and Intuition (N) types in Figure 7 (b) within Exp 1's unguided flying, differences in physiological reactions are notable in the PNS and SNS indices. Sensing individuals (colored in light blue and blue) show wider variances (PNS Index (0.31, std 0.44) and SNS Index (0.53, std 0.38)), reflecting immediate and tangible engagement. In contrast, Intuitive types (in red and pink) portray narrower variances (PNS (0.19, std 0.28) and SNS (0.42, std 0.34),), denoting a forward-thinking, exploratory tact.

During the heat detection process of Exp 2, pronounced physiological responses register across all indices, with Sensing types expressing consistent, detailed responses in SD1 and SD2, implying a systematic, intricate examination focus. Conversely, Intuitive subjects indicate diverse reactions, alluding to a penchant for creative, conceptual assessments, with elevated stress levels detected.

In the post-survey relaxation phase once again reveals a tranquilization in physiological markers, heralding a homeward swing to baseline states. Both types experience a dip in SI, approaching zero or slightly negative. Sensors exhibit a slight positive Stress Index (0.05, std 0.25), while Intuitives go to –0.05 (std 0.22), indicative of varied relaxation rates.

### 4.2.3 Thinking (T) vs. Feeling (F)

In figure 7 (c), focusing on Thinking (T) versus Feeling (F) types during Exp 1's free flight, noticeable variations in physiological responses surface, particularly in the PNS and SNS indices. Thinkers (showcased in red and pink) report a PNS Index mean of 0.32 (std 0.41) and an SNS Index mean of 0.56 (std 0.31), signifying methodical and analytical management, whilst Feelers (illustrated in light blue and blue) show lower PNS (0.13, std 0.21) and SNS (0.33, std 0.37) indices, but a Stress Index around 0.22 (std 0.29), reveal broader swings, indicative of deeper emotional involvement and reactivity.

In Exp 2, during the thermal leakage detection task, all types show intensified physiological responses. Here, Thinkers register structured and uniform responses across SD1 and SD2, indicating a logical, orderly approach, while Feelers exhibit wider response spectra hinting at an empathetic, values-oriented disposition and higher perceived stress with Stress Index (0.02, std 0.29).

The subsequent post-survey relaxation phase indicates physiologic measures decrease across both groups. Thinkers show SI at –0.03 (std 0.21), while Feelers hover around 0.00 (std 0.26), indicating slightly longer recovery times for Feelers.

**Table 4. Comparison of Heat Loss Detection Performance with MBTI**

| MBTI | Dimension | Physiological Factors | Exp 1 | | Exp 2 | | Relaxation | |
|---|---|---|---|---|---|---|---|---|
| | | | mean | std | mean | std | mean | std |
| E/I | E | PNS Index | 0.24 | 0.39 | 0.21 | 0.50 | -0.03 | 0.24 |
| | | SNS Index | 0.39 | 0.35 | 0.39 | 0.70 | -0.10 | 0.27 |
| | | SD1 | -0.15 | 0.22 | -0.11 | 0.36 | 0.09 | 0.27 |
| | | SD2 | -0.18 | 0.25 | -0.18 | 0.24 | 0.04 | 0.26 |
| | | Stress Index (SI) | 0.21 | 0.23 | 0.19 | 0.36 | -0.05 | 0.23 |
| | I | PNS Index | 0.24 | 0.30 | 0.04 | 0.29 | -0.15 | 0.21 |
| | | SNS Index | 0.55 | 0.34 | 0.05 | 0.36 | -0.10 | 0.30 |
| | | SD1 | -0.23 | 0.18 | -0.05 | 0.12 | 0.07 | 0.30 |
| | | SD2 | -0.27 | 0.19 | -0.01 | 0.22 | -0.08 | 0.26 |
| | | Stress Index (SI) | 0.34 | 0.24 | 0.05 | 0.21 | 0.03 | 0.23 |
| N/S | N | PNS Index | 0.19 | 0.28 | 0.02 | 0.31 | -0.13 | 0.24 |
| | | SNS Index | 0.42 | 0.34 | 0.08 | 0.55 | -0.17 | 0.26 |
| | | SD1 | -0.15 | 0.24 | -0.01 | 0.31 | 0.12 | 0.32 |
| | | SD2 | -0.22 | 0.25 | -0.07 | 0.25 | 0.01 | 0.31 |
| | | Stress Index (SI) | 0.24 | 0.22 | 0.05 | 0.33 | -0.05 | 0.22 |
| | S | PNS Index | 0.31 | 0.44 | 0.32 | 0.52 | -0.01 | 0.21 |
| | | SNS Index | 0.53 | 0.38 | 0.52 | 0.59 | 0.02 | 0.28 |
| | | SD1 | -0.23 | 0.13 | -0.21 | 0.20 | 0.01 | 0.17 |
| | | SD2 | -0.22 | 0.18 | -0.17 | 0.24 | -0.04 | 0.16 |
| | | Stress Index (SI) | 0.30 | 0.27 | 0.26 | 0.24 | 0.05 | 0.25 |
| F/T | F | PNS Index | 0.13 | 0.21 | 0.03 | 0.31 | -0.04 | 0.24 |
| | | SNS Index | 0.33 | 0.37 | 0.04 | 0.50 | -0.06 | 0.31 |
| | | SD1 | -0.13 | 0.26 | 0.03 | 0.35 | 0.07 | 0.31 |
| | | SD2 | -0.19 | 0.28 | -0.06 | 0.24 | -0.02 | 0.28 |
| | | Stress Index (SI) | 0.22 | 0.29 | 0.02 | 0.29 | 0.00 | 0.26 |
| | T | PNS Index | 0.32 | 0.41 | 0.22 | 0.49 | -0.11 | 0.23 |
| | | SNS Index | 0.56 | 0.31 | 0.40 | 0.63 | -0.12 | 0.25 |
| | | SD1 | -0.22 | 0.15 | -0.17 | 0.20 | 0.09 | 0.26 |
| | | SD2 | -0.24 | 0.18 | -0.15 | 0.25 | 0.00 | 0.26 |
| | | Stress Index (SI) | 0.30 | 0.19 | 0.22 | 0.30 | -0.03 | 0.21 |
| P/J | J | PNS Index | 0.20 | 0.27 | 0.06 | 0.29 | -0.06 | 0.22 |
| | | SNS Index | 0.36 | 0.34 | 0.16 | 0.59 | -0.09 | 0.29 |
| | | SD1 | -0.13 | 0.20 | -0.06 | 0.22 | 0.05 | 0.27 |
| | | SD2 | -0.18 | 0.23 | -0.11 | 0.22 | -0.02 | 0.24 |
| | | Stress Index (SI) | 0.20 | 0.25 | 0.10 | 0.29 | -0.03 | 0.25 |
| | P | PNS Index | 0.30 | 0.24 | 0.26 | 0.14 | -0.11 | -0.08 |
| | | SNS Index | 0.63 | 0.46 | 0.38 | 0.25 | -0.11 | -0.10 |
| | | SD1 | -0.27 | -0.18 | -0.13 | -0.08 | 0.12 | 0.08 |
| | | SD2 | -0.29 | -0.22 | -0.12 | -0.11 | 0.02 | -0.01 |
| | | Stress Index (SI) | 0.37 | 0.26 | 0.19 | 0.13 | 0.00 | -0.02 |

### 4.2.4 Judging (J) vs. Perceiving (P)

For Judging (J) and Perceiving (P) personalities in Figure 7 (d), Exp 1's free-flight stage exhibits considerable variability in physiological measures, especially within PNS and SNS indices. Distinct patterns emerge for Judging (light blue and blue) and Perceiving (red and pink) individuals have higher values for PNS (0.30, std 0.24) and SNS (0.63, std 0.46) - with a notably higher range observed among Perceiving types.

During Exp 2's structured heat loss identification task, both types show moderated physiological responses, though important differences persist. Perceiving types maintain higher PNS (0.26, std 0.14) and SNS indices (0.38, std 0.25) compared to Judging types (PNS: 0.06, std 0.29; SNS: 0.16, std 0.59), suggesting sustained engagement. However, the gap narrows considerably, indicating Judging types' improved physiological comfort during structured tasks. Notably, Judging types display greater variability (indicated by higher standard deviations across metrics), suggesting more diverse response patterns when confronted with well-defined objectives. Perceiving types maintain higher stress indices (0.19, std 0.13 vs. J: 0.10, std 0.29) but with reduced variability, indicating more consistent stress responses during targeted detection tasks.

In the post-relaxation stage, both types return toward baseline, though with different recovery patterns. Judging types show a more pronounced parasympathetic rebound (PNS: -0.06, std 0.22) compared to Perceiving types (PNS: -0.11, std -0.08), suggesting potentially more efficient recovery. The stress index for Judging types drops below baseline (SI: -0.03, std 0.25), while Perceiving types return precisely to baseline (SI: 0.00, std -0.02). SD1 values show interesting contrasts, with Judging types averaging 0.05 (std 0.27) versus Perceiving types' 0.12 (std 0.08), potentially indicating different vagal recovery mechanisms.

### 4.2.5 Integrated Observations

Extraversion (E) generally correlates with more pronounced SNS and Stress Index changes during Exp 1. free flight and Exp 2. heat loss detection, whereas Introversion (I) aligns with steadier but sometimes slightly elevated stress indicators in Exp 2. Sensors (S) consistently show higher engagement metrics across Exp 2 and 3, while Intuitives (N) exhibit broader fluctuations in SD1 and SD2, reflecting conceptual exploration and variable stress responses. Thinkers (T) maintain more uniform SD1 and SD2 patterns, suggesting methodical coping strategies, while Feelers (F) demonstrate wider stress index fluctuations, underscoring empathetic, emotion-driven behavior. Judgers (J) maintain tighter physiological ranges and stable stress profiles, in contrast to Perceivers (P), whose adaptability manifests as higher variance in PNS, SNS, and SI, especially during complex tasks.

### 4.3 Heat Loss Detection Performance Analysis

To evaluate the effectiveness of VR-based training for drone-assisted heat loss detection, we analyzed participants' performance based on their MBTI personality types. The key performance indicators included accuracy, recall, precision, F1-score, false rate of wrongly taken images, and coverage rate, allowing for a comprehensive assessment of their ability to identify heat loss areas correctly. Tables 5 summarize these findings by MBTI personality types, and Figures 8 and 9 provide a visual comparison of performance.

We analyzed the frequency and distribution of different error types across MBTI personality traits to assess participants' misclassification patterns in drone-assisted heat loss detection further. The primary error categories include misidentifying air conditioning (AC) units as heat loss, analyzing the wrong building, capturing areas with no relevant information (regular roof/wall), capturing the sky, and capturing the floor. The total number of false detections per participant was also recorded under the FALSE Summary column, representing the overall misclassification tendency. The statistical results are presented in Table 6.

**Table 5. Comparison of Heat Loss Detection Performance with MBTI**

| MBTI | Dimension | Accuracy | | Recall | | Precision | | F1 | | False Rate of Wrongly Taken Images | | Coverage Rate | |
|---|---|---|---|---|---|---|---|---|---|---|---|---|---|
| | | mean | std | mean | std | mean | std | mean | std | mean | std | mean | std |
| E/I | E | 0.36 | 0.18 | 0.29 | 0.20 | 0.68 | 0.31 | 0.38 | 0.24 | 0.32 | 0.32 | 0.76 | 0.64 |
| | I | 0.28 | 0.17 | 0.21 | 0.19 | 0.57 | 0.29 | 0.30 | 0.22 | 0.38 | 0.29 | 0.66 | 0.56 |
| N/S | N | 0.39 | 0.16 | 0.32 | 0.19 | 0.71 | 0.25 | 0.42 | 0.22 | 0.26 | 0.23 | 0.91 | 0.65 |
| | S | 0.23 | 0.17 | 0.16 | 0.17 | 0.51 | 0.34 | 0.22 | 0.21 | 0.48 | 0.37 | 0.41 | 0.33 |
| F/T | F | 0.39 | 0.18 | 0.32 | 0.20 | 0.72 | 0.27 | 0.42 | 0.24 | 0.28 | 0.26 | 0.88 | 0.66 |
| | T | 0.24 | 0.13 | 0.17 | 0.14 | 0.52 | 0.31 | 0.24 | 0.19 | 0.44 | 0.35 | 0.50 | 0.42 |
| P/J | J | 0.33 | 0.17 | 0.27 | 0.19 | 0.71 | 0.27 | 0.37 | 0.23 | 0.32 | 0.28 | 0.87 | 0.59 |
| | P | 0.32 | 0.20 | 0.23 | 0.20 | 0.51 | 0.31 | 0.31 | 0.24 | 0.38 | 0.35 | 0.48 | 0.55 |

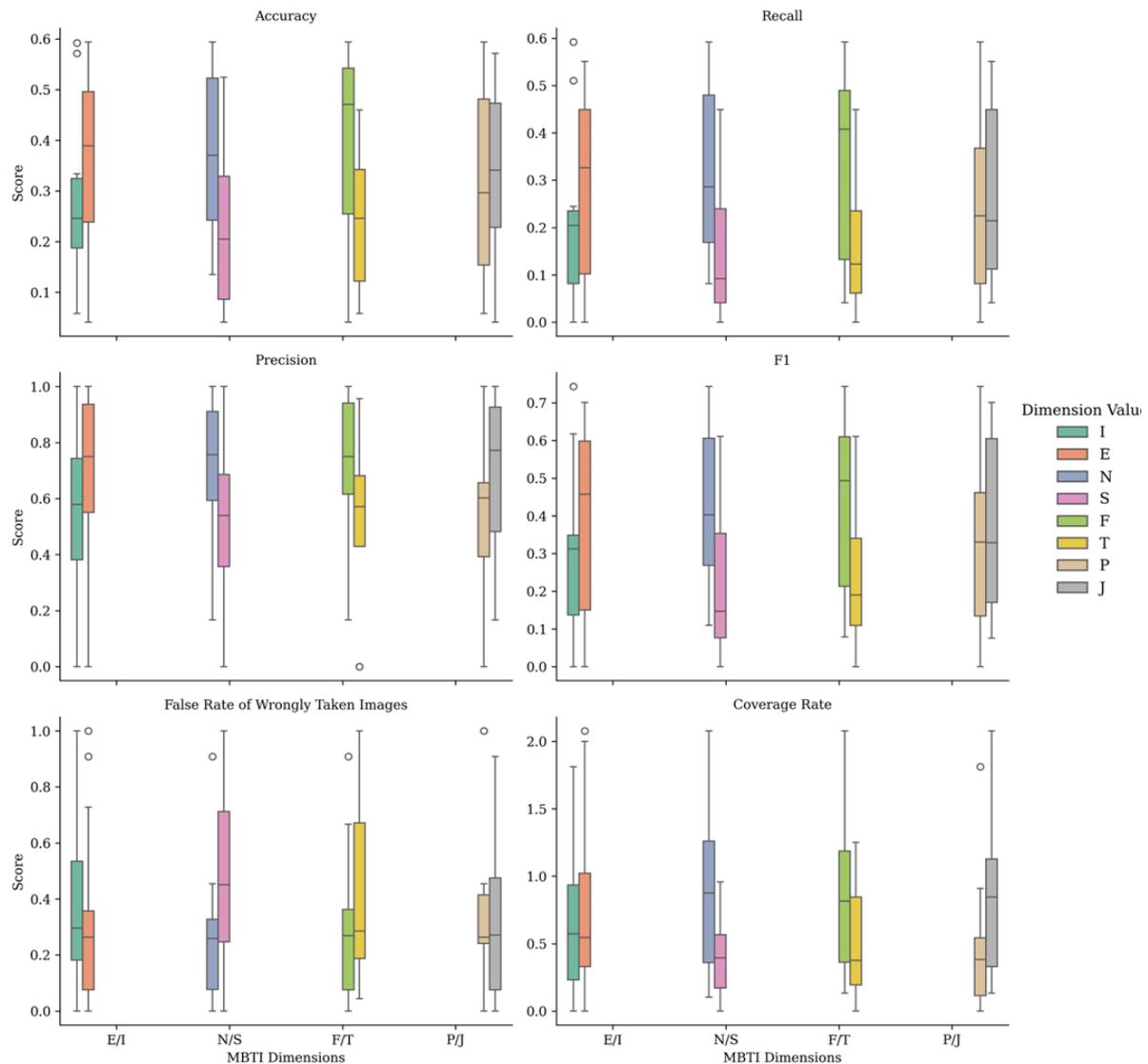

Figure 8. Comparison of Heat Loss Detection Performance with MBTI

**Table 6. Comparison of Misclassification Performance with MBTI**

| MBTI | Dimension | Treat AC as heat loss | | Wrong Building | | No info Regular Roof/Wall | | Sky | | Floor | | False Summary | |
|---|---|---|---|---|---|---|---|---|---|---|---|---|---|
| | | mean | std | mean | std | mean | std | mean | std | mean | std | mean | std |
| E/I | E | 0.47 | 1.25 | 3.53 | 5.58 | 0.47 | 0.83 | 1.07 | 2.15 | 0.20 | 0.56 | 5.73 | 5.93 |
| | I | 1.27 | 2.10 | 1.73 | 2.00 | 1.00 | 1.48 | 0.55 | 1.29 | 0.36 | 1.21 | 4.91 | 2.84 |
| N/S | N | 1.06 | 1.88 | 2.31 | 4.92 | 1.13 | 1.31 | 0.94 | 2.11 | 0.13 | 0.50 | 5.56 | 5.50 |
| | S | 0.40 | 1.26 | 3.50 | 3.72 | 0.00 | 0.00 | 0.70 | 1.34 | 0.50 | 1.27 | 5.10 | 3.70 |
| F/T | F | 0.73 | 1.39 | 3.20 | 5.47 | 0.80 | 1.32 | 0.87 | 2.10 | 0.20 | 0.56 | 5.80 | 5.89 |
| | T | 0.91 | 2.07 | 2.18 | 2.68 | 0.55 | 0.93 | 0.82 | 1.47 | 0.36 | 1.21 | 4.82 | 2.93 |
| P/J | J | 0.44 | 1.50 | 2.63 | 5.23 | 0.56 | 0.89 | 0.69 | 1.30 | 0.38 | 1.09 | 4.69 | 5.03 |
| | P | 1.40 | 1.84 | 3.00 | 3.09 | 0.90 | 1.52 | 1.10 | 2.51 | 0.10 | 0.32 | 6.50 | 4.45 |

## 4.3.1 Extraversion (E) vs. Introversion (I)

As shown in Table 6, extraverted (E) participants demonstrated a higher mean accuracy (0.36) compared to introverted (I) participants (0.28). Similarly, recall values for extraverts (0.29) exceeded those of introverts (0.21), indicating that extraverts were more likely to detect heat loss areas but also exhibited a higher false positive rate. Additionally, extraverts showed a higher precision (0.68) than introverts (0.57), suggesting that while they identified more heat loss areas, their results accurately distinguished true heat loss from background noise. The false rate of wrongly taken images was slightly lower for extraverts (0.32) compared to introverts (0.38), while the coverage rate, representing the proportion of heat loss identified relative to total images taken, was also higher for extraverts (0.76) than introverts (0.66). This suggests that extraverts explore larger areas and take more images, improving their overall detection capability.

In the E/I (Extraversion vs. Introversion) dimension, extraverts (E) were more likely to select the wrong building (mean: 3.53) compared to introverts (I) (mean: 1.73), suggesting that extraverts tend to explore a broader area, occasionally leading to incorrect target selection. However, introverts were more prone to misclassifying AC units as heat loss (mean: 1.27 vs. 0.47 for extraverts), possibly due to a more cautious inspection approach that misinterprets heat-emitting objects. In contrast, extraverts had a higher false rate for capturing the sky (mean: 1.07 vs. 0.55), indicating that their frequent aerial maneuvering may result in unintended captures. Extraverts had a slightly higher FALSE Summary (5.73) than introverts (4.91), reflecting their broader but sometimes imprecise exploration patterns.

## 4.3.2 Sensing (S) vs. Intuition (N)

When considering the N/S (Intuition vs. Sensing) dichotomy, intuitive (N) participants significantly outperformed sensing (S) participants across all key metrics. Accuracy for intuitive individuals was recorded at 0.39, whereas sensing participants achieved only 0.23. Similarly, intuitive participants exhibited superior recall (0.32) and precision (0.71), leading to a higher F1-score (0.42) compared to sensing participants (0.22). The false rate of wrongly taken images was substantially lower for intuitive participants (0.26) than for sensing participants (0.48), indicating that sensing participants were more prone to misclassification errors, such as mistaking non-heat-loss areas for heat loss. Additionally, intuitive participants had a much higher coverage rate (0.91) than sensing participants (0.41), suggesting that their exploration and pattern-recognition tendencies allowed them to detect more heat loss areas comprehensively.

For the N/S (Intuition vs. Sensing) dimension, intuitive (N) participants demonstrated a more balanced error distribution across all categories. They showed a higher misclassification rate for AC units (1.06) and regular roof/wall areas (1.13) compared to sensing (S) participants, who had almost no errors (0.00) in the regular roof/wall category. However, sensing participants had a significantly higher error rate for selecting the wrong building (3.50) compared to intuitive participants (2.31), indicating that sensors were more likely to misidentify their intended inspection target. The FALSE Summary for intuitive participants (5.56) and sensing participants (5.10) suggests that intuitive participants tend to make diverse errors across categories, whereas sensing participants exhibit a higher frequency of specific errors.

### 4.3.3 Thinking (T) vs. Feeling (F)

A similar trend was observed in the F/T (Feeling vs. Thinking) dimension. Feeling (F) participants consistently outperformed thinking (T) participants in accuracy (0.39 vs. 0.24), recall (0.32 vs. 0.17), and precision (0.72 vs. 0.52). The F1-score for feeling participants was 0.42, whereas thinking participants had only 0.24, highlighting the greater ability of feeling participants to balance recall and precision effectively. The false rate of wrongly taken images was lower for feeling participants (0.28) compared to thinking participants (0.44), indicating a more careful and intuitive approach to image selection. Additionally, the coverage rate for feeling participants (0.88) was notably higher than for thinking participants (0.50), suggesting that their inspection strategy was more effective in covering heat loss areas while minimizing errors.

When comparing F/T (Feeling vs. Thinking) personalities, feeling (F) participants exhibited a higher tendency to incorrectly classify heat loss in non-relevant areas, particularly in regular roof/wall sections (0.80) and wrong buildings (3.20). On the other hand, thinking (T) participants showed a slightly higher error rate for AC misclassification (0.91 vs. 0.73 for feeling participants), indicating a potential bias toward over-relying on heat signatures without contextual verification. Both groups exhibited similar error rates in capturing the sky and the floor, suggesting that this mistake is not strongly influenced by personality type. However, the overall misclassification rate (FALSE Summary) was slightly higher for feeling participants (5.80) than thinking participants (4.82), suggesting that feeling participants might be more prone to errors in complex decision-making scenarios.

### 4.3.4 Judging (J) vs. Perceiving (P)

Lastly, the P/J (Perceiving vs. Judging) dimension revealed that judging (J) participants exhibited slightly better overall performance than perceiving (P) participants. Judging participants achieved higher accuracy (0.33 vs. 0.32), recall (0.27 vs. 0.23), and precision (0.71 vs. 0.51). Their false rate of wrongly taken images was lower (0.32) than that of perceiving participants (0.38), suggesting they were more systematic and deliberate in their inspection process. The coverage rate for judging participants (0.87) was also significantly higher than for perceiving participants (0.48), implying that a structured and methodical approach led to more thorough and accurate heat loss detection.

Lastly, in the P/J (Perceiving vs. Judging) dimension, perceiving (P) participants exhibited significantly higher false rates across most categories compared to judging (J) participants. They had a particularly high rate of AC misclassification (1.40), incorrect building selection (3.00), and regular roof/wall misclassification (0.90), indicating that perceiving participants may struggle with maintaining structured search patterns. Additionally, perceiving participants had a notably higher overall false detection rate (FALSE Summary: 6.50) compared to judging participants (4.69). These results suggest that judging participants' more methodical approach led to fewer misclassification errors, whereas perceiving participants as being more flexible and exploratory were more prone to mistakes

### 4.3.5 Integrated Observations

Extraverts and Intuitives excel at locating heat-loss areas quickly (high recall/coverage) but can incur more off-target captures (wrong building, sky). Sensors often misidentify building targets, while Intuitives occasionally mistake non-heat-loss elements (e.g., AC units) for true defects. Feelers demonstrate strong detection metrics but sometimes commit more errors in ambiguous scenes, whereas Thinkers rely on logic and can miss subtle cues. Judgers systematically minimize random errors, while Perceivers explore broadly at the cost of higher misclassification rates. Extraverts, intuitive, and perceiving participants demonstrated broader exploration tendencies, leading to higher rates of incorrect target selection. In contrast, introverts, sensing, and judging participants showed more focused but occasionally overly cautious behaviours, leading to misclassifications of AC units and false positives in heat loss detection.

In sum, personality traits significantly impact drone-based heat loss detection. Participants with greater openness, intuition, and feeling-oriented approaches tend to perform better across most metrics, while Extraverts and Judgers achieve higher overall task efficiency. These insights suggest that tailoring VR training to individual MBTI profiles—e.g., providing structured flight plans for Perceivers or cautionary prompts for Extraverts—may further enhance the accuracy and reliability of drone-based inspections.

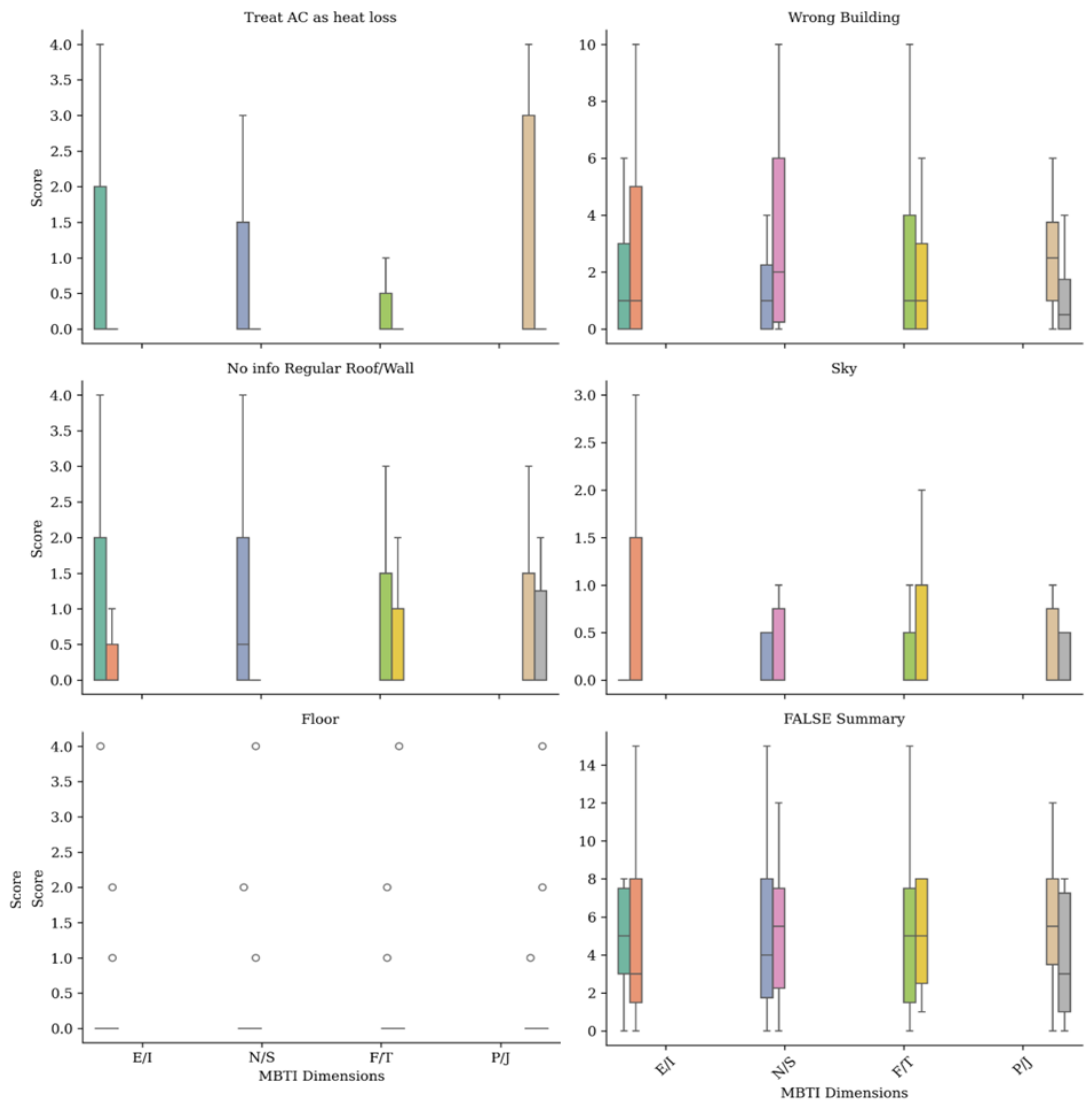

Figure 9. Comparison of Misclassification Performance with MBTI

## 4.4 Limitations

While this study provides valuable insights into the impact of MBTI personality traits on drone-assisted heat loss detection performance in a VR-based learning environment, several limitations should be acknowledged. First, VR-based simulations differ from real-world drone operations, particularly in terms of environmental complexity, physical constraints, and operational stress. In a controlled VR setting, participants do not experience external factors such as wind interference, battery limitations, real-time risk assessment, or hardware malfunctions, which could significantly affect drone operation and decision-making in real-world inspections. The absence of haptic feedback and tactile interaction may also influence learning retention and task execution strategies. Future research should explore mixed-reality or real-world validation studies to bridge the gap between simulated training and onsite drone operations. Second, while the study analyzed physiological responses using HRV metrics, additional neurophysiological and behavioral data could enhance our understanding of cognitive workload, stress adaptation, and decision-making patterns. The integration of eye-tracking technology, electroencephalography (EEG), or galvanic skin response (GSR) could provide a deeper analysis of visual attention, mental engagement, and cognitive stress during drone operations. Additionally, the study focused on MBTI personality traits, which, while widely used, may not fully capture all cognitive and behavioral factors affecting inspection performance. Other personality models, such as the Big Five Personality Traits (OCEAN model) or cognitive style assessments, could provide a more nuanced understanding of individual differences in perception, decision-making, and adaptability in a VR-based drone training environment. Lastly, task complexity and environmental diversity in the VR simulation were relatively controlled, which may not fully reflect the varied scenarios encountered in real-world heat loss inspections. Future studies should introduce dynamic weather conditions, variable lighting, and additional structural complexity to simulate realistic challenges that drone operators face in the field. Furthermore, longitudinal studies examining learning retention over time could determine whether VR-based training effectively translates into long-term skill acquisition and operational proficiency in real-world inspections.

Despite these limitations, the findings provide a foundation for improving VR-based workforce training for drone-assisted energy audits and infrastructure inspections. Addressing these challenges in future research could enhance learning effectiveness, training adaptability, and real-world applicability, ultimately improving the efficiency and accuracy of drone-based building diagnostics.

## 5. Conclusion and Future Work

This study provides valuable insights into how the MBTI factor influences inspection performance in VR-based drone-assisted heat loss detection, emphasizing three critical dimensions: flight trajectory patterns, stress adaptation, and task performance. Key findings reveal that:

1. **Flight Path Strategies**: Personality traits significantly influenced drone operation strategies. Extraverted (E) participants adopted broader exploration patterns, covering more area than introverts (I), who tended to adopt a more cautious and focused approach. Introverts focused on localized inspections, reducing spatial errors but increasing misclassifications. Similarly, sensing (S) participants followed linear and clustered flight paths, while intuitive (N) participants exhibited more dispersed and exploratory movements. Thinking (T) maintained systematic, logical trajectories, while Feeling (F) showcased flexible and empathetic movement. Judging (J) participants maintained structured and predictable navigation, whereas perceiving (P) participants adjusted their trajectories dynamically. These results highlight the importance of designing training modules that accommodate different learning styles, ensuring that both exploratory and structured learners benefit from VR-based workforce training.

2. **Stress Level and Adaptation**: The physiological analysis revealed that participants initially experienced elevated stress levels, especially during their first interactions with the VR-based drone control system. However, stress levels decreased over time as they became more familiar with the environment and task complexity. Analysis revealed that initial stress levels (Stress Index) decreased after repeated training sessions, indicating improved cognitive adaptation to VR environments. HRV metrics (PNS, SNS, SD1, SD2, and Stress Index) indicated that Extraverts, Intuitives, Feelers, and Perceivers generally experienced broader stress fluctuations under complex inspection tasks. Introverts, Sensors, Thinkers, and Judgers typically displayed more stable physiological responses, suggesting a preference for structured methods and steady-state coping strategies. Task complexity (e.g., heat loss detection) amplified existing personality-based differences, highlighting how varying cognitive and emotional styles affect performance under pressure.

3. **Performance in Heat Loss Detection**: Performance evaluation metrics, including accuracy, recall, precision, and error types, showed that intuitive (N) and feeling (F) participants outperformed others in heat loss detection , while judges (J) achieved higher precision and coverage rates due to methodical workflows. Extraverts (E) demonstrated higher exploration efficiency but also a higher false detection rate, whereas introverts (I) exhibited more conservative and targeted inspection behaviors. Error analysis highlighted critical training gaps, such as misinterpreting thermal signatures (errors involved AC units) and operational misalignment (e.g., sky/floor captures). Extraverts, Intuitives, and Feelers often excelled in recall and coverage rate, suggesting strong exploratory and empathetic tendencies. However, they also demonstrated higher error rates in certain contexts, such as misclassifying or over capturing non-defective areas. Introverts, Sensors, Thinkers, and Judgers tended to produce fewer random errors and overall more systematic approaches. Still, they occasionally missed subtle indications of heat loss or demonstrated overly cautious inspection strategies.

Given the increasing role of drones in energy audits and infrastructure inspections, future research should explore more advanced VR-based workforce training frameworks. Specifically, several areas warrant further investigation: (1) Personalized Training Modules: Future workforce training programs should leverage AI-driven adaptive learning techniques to tailor training difficulty based on individual learning progress, stress response, and performance trends. (2) MR Integration: To bridge the gap between VR simulations and real-world inspections, incorporating MR technologies could enhance realism, provide haptic feedback, and simulate real-world environmental challenges such as wind, lighting variations, and drone battery limitations. (3) Longitudinal Learning Retention Studies: Future work should explore how VR-based training influences long-term knowledge retention and skill transfer in real-world drone operations, assessing whether trainees maintain efficiency and accuracy over time. (4) Neurophysiological and Cognitive Assessments: Expanding the scope of cognitive workload analysis through EEG-based studies, eye-tracking, and real-time stress monitoring could provide deeper insights into trainee decision-making processes and visual attention.

## Acknowledgments

The authors would like to express their gratitude for the financial support provided by the WNE Alumni Association. The opinions expressed in this paper are solely those of the authors.The authors would like to express their gratitude for the financial support provided by the WNE Alumni Association. The opinions expressed in this paper are solely those of the authors.

# Appendix

### Pre-Survey (VR and AR)

| Do you have a mental disease? | Yes, No, I dont know |
| --- | --- |
| Have you ever experienced AR or VR content? | Yes, No, I dont know |
| Have you ever experienced 3D motion sickness? | Yes, No, I dont know |
| Did you sleep well for more than six hours? | Yes, No, I dont know |
| Did you drink coffee within 24 h? | Yes, No, I dont know |
| Did you drink within 24 h? | Yes, No, I dont know |
| Did you smoke within 24 h? | Yes, No, I dont know |
| Did you feel dizzy when you took the airplane? | Yes, No, I dont know |
| Did you feel dizzy when you took the ground vehicles, such as cars, buses, or trains? | Yes, No, I dont know |
| Did you feel dizzy when you took water transportation, such as a canoe, raft, submarine, surfboard, sailboat, or steamboat? | Yes, No, I dont know |
| How do you evaluate training using VR? | 1 to 5 (good) |
| How was the VR training compared to traditional training? | 1 to 5 (good) |
| Evaluate your physical condition. | 1 to 5 (good) |
| Evaluate your mental condition. | 1 to 5 (good) |

### Post Survey (VR and AR)

| To what extent did the game hold your attention? | 1 to 5 (high) |
| --- | --- |
| Evaluate the playing time (time) | 1 to 5 (long) |
| Evaluate how you feel about this application (program) | 1 to 5 (excited) |
| Evaluate the comfort of surroundings (environment) | 1 to 5 (good) |
| Were you interested in the application (interest)? | 1 to 5 (interested) |
| Was the application difficult (difficulty)? | 1 to 5 (easy) |
| Evaluate the immersiveness of the application (immersion). | 1 to 5 (high) |
| Evaluate the ability to control the drone (control). | 1 to 5 (high) |
| To what extent did you feel emotionally attached to the game? | 1 to 5 (high) |
| To what extent did you enjoy the graphics and the imagery? | 1 to 5 (high) |
| To what extent was your arm tired after playing the game? | 1 to 5 (tired) |
| To what extent were your eyes tired after playing the game? | 1 to 5 (tired) |
| Did you experience nausea or motion sickness during the training, and if so, when? | Yes, No, I dont know |
| Would you like to play the game again? | Yes, No, I dont know |
| If so, when? | Open questions |
| What are the advantages and disadvantages of training with the use of VR? | Open questions |
| Please predict your performance. | 1 to 5 (high) |

**Demographic Information Survey**

- **What is your age?**
  - Under 18
  - 18-24 years old
  - 25-34 years old
  - 35-44 years old
  - 45-54 years old
  - 55-64 years old
  - Overs
- **Select your gender**
  - Male
  - Female
  - Non-binary/third gender
  - Prefer not to say
- **Please select your ethnicity**
  - White
  - Hispanic/ Latino
  - Black or African American
  - American Indian or Alaska Native

- ○ Asian
- ○ Native Hawaiian or Pacific Islander
- ○ Other
● **Please select your level of education**
    - ○ Less than high school
    - ○ High school graduate
    - ○ Some college
    - ○ 2-year degree
    - ○ 4-year degree
    - ○ Professional Degree
    - ○ Masters
    - ○ Doctorate
    - ○ others
● **What is your current employment status?**
    - ○ Employed full-time (40+ hours a week)
    - ○ Employed part-time (less than 40 hours a week)
    - ○ Unemployed (currently looking for work)
    - ○ Unemployed (not currently looking for work)
    - ○ Student
    - ○ Retired
    - ○ Self-employed
    - ○ Others
● **If you are a current college student, what's your major?**
    - ○ Civil, Construction, and Environmental Engineering
    - ○ Electrical Engineering
    - ○ Computer Science/ Engineering
    - ○ Industrial Engineering/ Engineering Management
    - ○ Biomedical and Chemical Engineering
    - ○ Mechanical and Aerospace Engineering
    - ○ Others belong to STEM (science, technology, engineering, and mathematics)
    - ○ Others, not STEM
● **If you are not a college student, which one can describe your current job?**
    - ○ Civil, Construction, and Environmental Engineering
    - ○ Electrical Engineering
    - ○ Computer Science/ Engineering
    - ○ Industrial Engineering/ Engineering Management
    - ○ Biomedical and Chemical Engineering
    - ○ Mechanical and Aerospace Engineering
    - ○ Others belong to STEM (science, technology, engineering, and mathematics)
    - ○ Others, not STEM
● **In the last five years, where do you live? (see the map below)**
    - ○ 1
    - ○ 2
    - ○ 3
    - ○ 4
    - ○ 5
    - ○ 6
    - ○ 7
    - ○ 8
● **Where did you live before you were 18 years old? (see the map below)**

- - 1
  - 2
  - 3
  - 4
  - 5
  - 6
  - 7
  - 8
- **What is your marital status?**
  - Single (never married)
  - Married
  - Inadomestic partnership
  - Divorced
  - Widowed
  - Others
- **Household income**
  - Below 10k
  - 10k-50k
  - 50k-100k
  - 100k-150k
  - 150k-200k
  - 200k-250k
  - Over 250k
  - Prefer not to say